\documentclass[10pt,a4paper,twocolumn,english,prb,aps,showpacs,floatfix,groupedaddress,superscriptaddress]{revtex4-1}

\usepackage{graphicx}
\usepackage{graphicx}
\usepackage{epsfig}
\usepackage[english]{babel}
\usepackage{amsmath}
\usepackage{amssymb}
\usepackage{amsfonts}
\usepackage{longtable}
\usepackage{dcolumn}
\usepackage{bm}
\usepackage{bbm}
\usepackage{nicefrac}
\usepackage{color,array}
\usepackage{colortbl}
\usepackage{dsfont}
\usepackage{mathtools}
\usepackage{subfigure}  
\usepackage{soul}
\usepackage{xcolor,cancel}
\usepackage{cleveref}

\def\bra#1{\mathinner{\langle{#1}|}}           
\def\ket#1{\mathinner{|{#1}\rangle}}           

\def\braket#1#2{\langle#1|#2\rangle}

\begin{document}

\newcommand{\be}{\begin{equation}}
\newcommand{\ee}{\end{equation}}
\newcommand{\bearr}{\begin{eqnarray}}
\newcommand{\eearr}{\end{eqnarray}}
\newcommand{\bseq}{\begin{subequations}}
\newcommand{\eseq}{\end{subequations}}
\newcommand{\nn}{\nonumber}
\newcommand{\reqn}{\eqref}
\newcommand{\pdag}{{\phantom{\dagger}}}
\newcommand{\vpdag}{{\vphantom{\dagger}}}
\newcommand{\talpha}{\widetilde{\alpha}}
\newcommand{\mac}{\mathcal}

\newcommand{\redd}[1]{\textcolor{red}{#1}}

\title{Massive spinons in  $S=1/2$ spin chains: spinon-pair operator representation}

\author{Mohsen Hafez-Torbati}
\email{mohsen.hafez@tu-dortmund.de}
\affiliation{Lehrstuhl f\"ur Theoretische Physik I, Technische Universit\"at Dortmund,
Otto-Hahn-Stra\ss e 4, 44221 Dortmund, Germany}

\author{G\"otz S.\ Uhrig}
\email{goetz.uhrig@tu-dortmund.de}
\affiliation{Lehrstuhl f\"ur Theoretische Physik I, Technische Universit\"at Dortmund,
Otto-Hahn-Stra\ss e 4, 44221 Dortmund, Germany}

\date{\rm\today}

\begin{abstract}
Spinons are among the generic excitations in one-dimensional spin systems; they
can be massless or massive. The quantitative description of massive spinons
poses a considerable challenge in spite of various variational approaches.
We show that a representation in terms of hopping 
and Bogoliubov spinon processes, which we call ``spinon-pair'' operators, 
and their combination is possible. We refer to such a representation as 
second quantized form. Neglecting
terms which change the number of spinons yields the variational results. 
Treating the bilinear and quartic terms by continuous unitary transformations
leads to considerably improved results. Thus, we provide the proof-of-principle
that systems displaying massive spinons as elementary excitations 
can be treated in second quantization based on spinon-pair representation.
\end{abstract}



\pacs{75.10.Pq, 75.10.Jm, 75.10.Kt, 71.10.Li, 02.30.Mv}

\maketitle

\section{Introduction}

Quantum magnets constitute a flourishing field of research. 
In particular, the search for unconventional excitations and their 
quantitative understanding represents an important issue of current interest.

The elementary excitations which are known best are spin waves or magnons.
They appear as massless Goldstone bosons in long-ranged ordered magnets
such as ferromagnets or antiferromagnets. Their effect of the total spin
of the system is integer, i.e., they change the total magnetization or the
sublattice magnetization by one $\hbar$ \cite{holst40,auerb94}. 
Henceforth, we set Planck's constant to unity for the sake of simplicity.
Another class of integer excitations are triplons, i.e., gapped dressed particles with
$S=1$ as they appear in valence bond solids, for instance in all models 
resulting from coupling spin dimers of $S=1/2$ in one dimension, see e.g.\  
Refs.\ \onlinecite{Sachdev1990,uhrig96b,Knetter2000,trebs00}, 
in two dimensions, see e.g.\ Refs.\ 
\onlinecite{Sachdev1990,sachd99,weiho99b,Uhrig2004},
and in three dimensions, see e.g.\ Refs.\ 
\onlinecite{takat97,matsu02,nohad04,sirke05,jense11}.

But in particular in low-dimensional systems, fractionalization may occur.
This means that the integer excitations decompose into several, mostly
two fractional excitations. The famous example are the $S=1/2$ spinons
in the nearest-neighbor Heisenberg chain 
\cite{Faddeev1981,Karbach1997,caux06,mouri13} with the Hamiltonian 
\be 
\label{eq:j1j2}
H:=J_1^\vpdag\sum_i \bm{S}_i^\vpdag \cdot \bm{S}_{i+1}^\vpdag 
+ J_2^\vpdag \sum_i \bm{S}_i^\vpdag \cdot \bm{S}_{i+2}^\vpdag,
\ee 
where $\bm{S}_{i}^\vpdag$ defines the spin $S=1/2$ operator at site $i$ and the sum 
runs over the  sites of a chain. 
The couplings $J_1^\vpdag$ and $J_2^\vpdag$ control the interaction strengths 
between nearest neighbor (NN) and next-nearest neighbor (NNN) sites, respectively.
 The NN case is given by $J_2=0$.
For later use, we define the relative coupling $\alpha:=J_2^\vpdag/J_1^\vpdag$ 
which is a measure of the degree of frustration. 

Another analytically solvable case of massless spinons is realized in the
Haldane-Shastry model \cite{halda88a,shast88}. The Hamiltonian
of this model is related to the one in Eq.\ \eqref{eq:j1j2}, but for certain long-range
couplings $J_n$. The concept of massless spinons
is used to develop effective or approximate descriptions of a multitude of systems, 
even if the microscopic Hamiltonian does not match perfectly \cite{ender10},
but it can provide a starting point for perturbative inclusion of inter-chain
couplings \cite{colde01a,bocqu01,kohno07}. 
Clearly, massless spinons represent an intensive field of research.

In addition, current research addresses massive spinons,
 i.e., spinons of which the creation requires a finite amount of energy.
In one dimension, strongly frustrated chains such as given
by the Hamiltonian \eqref{eq:j1j2} for $\alpha\gtrapprox 0.241$
display such elementary excitations \cite{Shastry1981,caspe84,Brehmer1998}.
Four-spin and six-spin interaction terms can be considered as well \cite{Tang2011,Tang2015} 
which lead to spontaneous dimerization even without frustration. 
In two dimensions, systems such as kagom\'e lattices are prone to be governed by
fractional massive spinons \cite{domma03,balen10,yan11a,depen12,norma16}
Even in three dimensions, fractionalization takes place leading to magnetic monopoles
\cite{caste08,morri09,kadow09}. But these occur in highly anisotropic spin models
which marks an important difference to the spinons mentioned above in one
and two dimensions.

Given the great interest in spinons and the fact that massive
spinons are less well understood than there massless counterparts
we study massive spinons in the present article.
We start from the description introduced by Shastry and Sutherland \cite{Shastry1981}
in a second quantized form. Conceptually, we extend the description of Shastry and 
Sutherland to general chains. As a proof-of-principle we will
consider the frustrated spin chain in \eqref{eq:j1j2} for arbitrary 
frustration $\alpha$. We do not attempt to define the creation 
or annihilation of single spinons. Instead, we introduce spinon-pair
operators which denote bilinear processes involving spinons, i.e., hopping
of spinons or pairwise creation or annihilation of them. In addition, we keep track
of the interactions of two spinons and of the decay of one spinon into three.
We show that it is indeed possible to systematically define 
the Hamiltonion in terms of spinon-pairs.

In a second step, we analyze the obtained second quantized Hamiltonian
by continuous unitary transformations to extract the physical relevant 
properties. The processes changing the number of spinons are rotated away
in this fashion. But their renormalizing effects on the physical properties
are retained, at least on the level of our approximations.
 The physical properties comprise the effective spinon dispersion
and the value of the spin gap in particular. In this way, we show
that a second quantized description of massive spinons is possible on
the proof-of-principle level. 

The results obtained are considerably improved over the variational
results. This illustrates the potential behind the idea to formulate
microscopic Hamiltonian in terms of their elementary excitations.
Of course, this route requires to know what these quantities are.
Often, however, this is the case. Thus, we believe that the
approach pursued here can be transferred to many other physical systems as well.

The article proceeds as follows. In Sect.\ \ref{sec:basis} we introduce
a complete spinon basis for the subspaces with total spin $S_{\rm t}=0$, 
$S_{\rm t}=\frac{1}{2}$, and $S_{\rm t}=1$. Subsequently, in 
Sect.\ \ref{sec:ortho} the
states in this basis are orthonormalized. This allows us to introduce spinon-pair operators
for second quantization in Sect.\ \ref{sec:pair-operators}. This formalism
is applied to the frustrated Heisenberg chain with nearest-neighbor and next-nearest neighbor
interactions in Sect.\ \ref{sec:chain}. The employed method of continuous unitary
transformations is introduced in Sect.\ \ref{sec:cut} while the final results are 
presented in Sect.\ \ref{sec:res}.
The conclusions and the outlook terminate the paper in Sect.\ \ref{sec:conclusio}.

\section{Spinon basis}
\label{sec:basis}

Here we introduce a basis for spinon states in chains consisting of localized spins $S=1/2$.
We distinguish between vacua and states with various numbers of spinons. This does not imply
that the vacua are ground states of the spin chains studied finally. The first aim of this
section and the subsequent two sections is to express the Hamiltonian in terms of spinons.
Then a continuous unitary transformation is applied to obtain quantitative results.

For a chain with an even number of sites, the 0-spinon state (spinon vacuum) is defined such 
that each spin at a site forms a singlet with its neighboring site. 
For a chain of length $L$, the 0-spinon state is given by  
\be
\label{eq:spinon0}
\ket{0}:=\!\!\!\bigotimes_{n=-L/4}^{L/4-1}[2n,2n+1] =: 
\includegraphics[trim=0.8 1.3 0.7 0,scale=10]{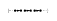}
\ee
where 
\be
[i,i+d]:=\includegraphics[trim=0.5 1.5 0.5 0,scale=10]{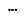}
:=\frac{1}{\sqrt{2}}\left( \ket{\uparrow}_{i}\ket{\downarrow}_{i+d} - 
\ket{\downarrow}_{i}\ket{\uparrow}_{i+d} \right) \quad
\ee
depicts the singlet state between lattice positions $i$ and $i+d$. The states 
$\ket{\uparrow}_{i}$ and $\ket{\downarrow}_{i}$ 
are eigenvectors of the $S^z_i$ operator with  eigenvalues $+1/2$ and $-1/2$, respectively. 
Planck's constant is set to unity.

For periodic boundary condition (PBC) there are two vacua which differ by a translation by one site.
For open boundary condition (OBC) these two states constitute the vacuum and a 2-spinon state
with spinons at both boundary points.
The spinon vacuum for OBC is a product of singlet; it
 is the same as the well-known Majumdar-Ghosh (MG) state which represents 
the exact ground state of the $J_1$-$J_2$ Heisenberg chain, see Eq.\ \reqn{eq:j1j2}, 
for $J_2=J_1/2 >0$ \cite{Majumdar1969a,Majumdar1969b}. 
This spinon vacuum can also be seen as a short-ranged ``resonating valence-bond'' (RVB) state 
\cite{Anderson1973,Oguchi1989} defined on a chain. Although
spinons as spin-$\frac{1}{2}$ quasiparticles always appear in pairs for given
even chain size, understanding the dynamics of a single spinon is necessary
to describe deconfined spinon pairs \cite{Shastry1981}. 
This parallels fermionic systems with a fixed number of particles. 

A spinon is defined as the domain-wall separating two possible vacua \cite{Shastry1981}. 
The 1-spinon state $\ket{\phi_{i}^{\sigma}}$ with the spin index $\sigma=\uparrow,\downarrow$ is given by 
\begin{align}
\ket{\phi_{i}^{\uparrow}}&:=\cdots \otimes[i-2,i-1]\otimes \uparrow_i^\vpdag \otimes [i+1,i+2] 
\otimes \cdots \nn \\
&=:\includegraphics[trim=0 1.2 0 0,scale=10]{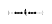}.
\label{eq:spinon1}
\end{align}
and similarly for $\ket{\phi_{i}^{\downarrow}}$. For OBC, the spinon can exist only on 
the odd or on the even sublattice, 
but for PBC it can be on either of them. 
The 1-spinon states are not orthonormal displaying the overlap \cite{Liang1988}
\be 
\label{eq:spinon1_ovlp}
\langle \phi_{i}^{\sigma} | \phi_{j}^{\sigma'} \rangle 
= \delta_{\sigma,\sigma'}\left(-\frac{1}{2} \right)^{\frac{|i-j|}{2}} 
\quad ; \quad (i-j) \in {\rm even},
\ee
and zero for $(i-j)$ odd in the limit $L \to \infty$.

The 2-spinon sector is spanned by singlet and triplet states. 
The singlet 2-spinon state $\ket{\phi_{i,i+d}^{s}}$ with $d\geq3$ 
reads
\be
\!\!\!\!
\ket{\phi_{i,i+d}^{s}}\!:=
\includegraphics[trim=1.5 1.3 1.2 1,scale=9.7]{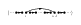} .
\label{eq:spinon2_s}
\ee
We notice that $\ket{\phi_{i,i+1}^{s}}=\ket{0}$ and hence the singlet 
2-spinon state with $d=1$ is not defined
because it is the vacuum state. In 
addition, there exists no 2-spinon state with even distance $d$.

Similarly, the triplet 2-spinon state $\ket{\phi_{i,i+d}^{t,p}}$ with  
$d\geq 1$ and the flavor $p=x,y,z$ is given by
\be
\hspace{-0.25cm}
\ket{\phi_{i,i+d}^{t,p}}:=
\includegraphics[trim=1.5 2.05 1.3 1,scale=9.6]{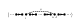} ,
\ee
where 
\begin{align}
\includegraphics[trim=1 1.6 0.8 0,scale=11]{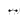}
&:= 
  \begin{cases}
    \frac{1}{\sqrt{2}}\left( \ket{\downarrow}_{j}\ket{\downarrow}_{j+d} - 
		\ket{\uparrow}_{j}\ket{\uparrow}_{j+d} \right), & p=x \\
    \frac{i}{\sqrt{2}}\left( \ket{\uparrow}_{j}\ket{\uparrow}_{j+d} + 
		\ket{\downarrow}_{j}\ket{\downarrow}_{j+d} \right), & p=y \\
    \frac{1}{\sqrt{2}}\left( \ket{\uparrow}_{j}\ket{\downarrow}_{j+d} + 
		\ket{\downarrow}_{j}\ket{\uparrow}_{j+d} \right), & p=z
  \end{cases}
\nn \\
&=: \{j,j+d\}_p^\vpdag
\end{align}
denotes a triplet bond with the flavor $p$ between sites $j$ and $j+d$ \cite{Sachdev1990}.  

The overlap between the spinon vacuum \reqn{eq:spinon0} and the singlet 
2-spinon state \reqn{eq:spinon2_s} is given by
\be 
\braket{0}{\phi_{i,i+d}^{s}}=\left( -\frac{1}{2} \right)^{\frac{d-1}{2}}.
\ee 
We also have the following overlap between singlet 2-spinon states
\be 
\braket{\phi_{j,j+d_2}^{s}}{\phi_{i,i+d_1}^{s}}=
\begin{cases}
 \left( -\frac{1}{2} \right)^{\frac{d_1+d_2}{2}-1} &,\, n > d_1  \\
 \left( -\frac{1}{2} \right)^{\frac{d_2-d_1}{2}+n} &,\,  d_1-d_2 \leq n < d_1  \\
 \left( -\frac{1}{2} \right)^{\frac{d_1-d_2}{2}} &,\, n \leq d_1-d_2  
\end{cases}
\ee
where $n:=j-i \geq 0$.  For triplet 2-spinon states,  the same relation holds,
except that for $n>d_1$ the result is zero.

How can we construct states which contain more than two spinons? 
In this manuscript, we focus on the subspaces with 
the total spin $S_{\rm tot}=0$ ($L$ even), $S_{\rm tot}=\frac{1}{2}$ ($L$ odd), and 
$S_{\rm tot}=1$ ($L$ even). 
For these cases, we show that a complete spinon basis can be  systematically constructed. 
The systematic construction is important for the second quantization process that we want 
to introduce in the sequel.

By introducing longer ranged singlets in the spinon vacuum we create various states with total spin zero. 
But these states are not all linearly independent. 
This is so because ``crossed'' singlets can be expressed in terms of 
``nested'' singlets and ``distinct'' singlets according to 
\be 
\includegraphics[trim=1 1.3 1 0.5,scale=9.2]{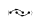} 
\!=\! \includegraphics[trim=1 1.3 1 0.5,scale=9.2]{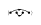} 
+ \includegraphics[trim=1 1.23 1 0.5,scale=9.2]{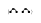} .
\label{eq:identity_sing}
\ee
This relation is valid for two arbitrary singlet bonds on arbitrary sites in the sequence of the chain.
By recursive application of Eq.\ \reqn{eq:identity_sing} any crossed singlet can be re-expressed in terms
of nested and distinct singlets. One can choose any two groups from the three groups
``distinct'', ``nested'', and ``crossed'' singlets to have a complete basis spanning the 
$S_{\rm tot}=0$ Hilbert space. Here, we adopt singlets of the
type ``distinct'' and ``nested'' as they are already used in the definition of the 
0-spinon and 2-spinon states.
Thus, this choice appears to be the most suitable to define spinon-pair operators.
We notice that the states with crossed singlets do not necessarily contain a well-defined number of spinons. 

Similar to the case of the $S_{\rm tot}=0$ subspace, by introducing singlet spinon pairs above 1-spinon states we can construct 3-spinon, 5-spinon, and higher spinon states with the total spin $S_{\rm tot}=\frac{1}{2}$.
The relation 
\be 
\label{eq:identity_spin_half}
\includegraphics[trim=0 18 0 0,scale=0.5]{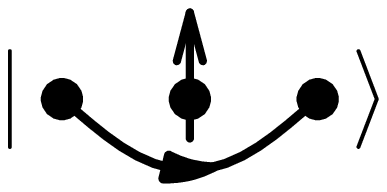}
= \includegraphics[trim=0 13 0 0,scale=0.5]{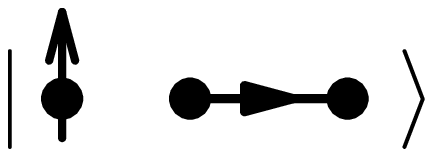} 
+\includegraphics[trim=0 13 0 0,scale=0.5]{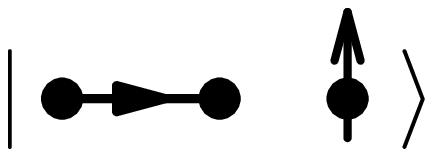}
\ee
makes the basis of all such states overcomplete. The  positions in \reqn{eq:identity_spin_half} 
are assumed to be arbitrary ones along the chain. Any state with a nested spinon can be expanded 
in terms of states in which the spinon is either  before or beyond the singlet bond.
In order to avoid overcounting, we restrict the $S_{\rm tot}=\frac{1}{2}$ subspace such that 
no single spinon occurs inside a singlet bond.

The states with $S_{\rm tot}=1$ can be generated by replacing one of the 
singlet bonds of a state in the $S_{\rm tot}=0$ subspace by a triplet bond. The identity
\be 
\includegraphics[trim=1 1.2 1 1,scale=9.2]{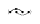} 
\!=\! \includegraphics[trim=1 1.4 1 0,scale=9.2]{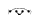} +
\includegraphics[trim=1 1.2 1 0,scale=9.2]{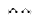} 
\label{eq:identity_trip_I}
\ee
together with Eq.\ \reqn{eq:identity_sing} justify why no crossed bond needs to be considered
in the $S_{\rm tot}=1$ subspace. 
In Eq.\ \reqn{eq:identity_trip_I} we omitted the flavor label $p$ from the triplet bonds 
because the relation is valid for any fixed value of $p$.
In addition to \reqn{eq:identity_trip_I}, we find 
\begin{align}
\label{eq:identity_trip_II}
\includegraphics[trim=1.8 1.15 1 0,scale=9.2]{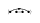} 
\!=\! \includegraphics[trim=1 1.2 1 0,scale=9.2]{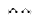} 
&+\includegraphics[trim=1 1.2 1 0,scale=9.2]{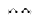} 
\nn \\
&+\includegraphics[trim=1 1.15 1 0,scale=9.2]{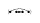} ,
\end{align}
which is a relation between distinct and nested singlet and triplet bonds. 
We notice that \reqn{eq:identity_trip_I} and \reqn{eq:identity_trip_II} are two independent equations. By recursive application of Eq.\ \reqn{eq:identity_trip_II} one can eliminate all
 nested triplet bonds. This means that in the construction of $S_{\rm tot}=1$ subspace, no 
nested triplet bond needs to be considered. 
Alternatively, one may decide to use Eq.\ \reqn{eq:identity_trip_II} to consider only 
nearest-neighbor (NN) triplet bonds in the $S_{\rm tot}=1$ subspace. But this leads to 
an overcomplete basis for the $S_{\rm tot}=1$ sector and further restriction is required 
which would complicate the subsequent treatment. 

\begin{figure}[thb]
  \centering
  \includegraphics[width=0.98\columnwidth,angle=0]{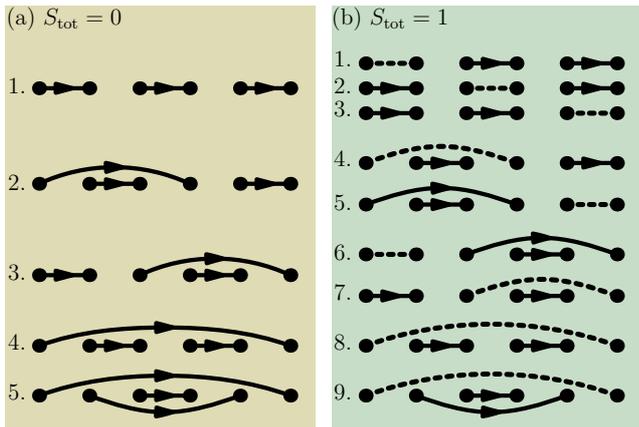}
  \caption{(Color online) The spinon states with (a) $S_{\rm tot}=0$  and (b)
  $S_{\rm tot}=1$ (b) on a 6-site cluster. Each triplet bond (dashed line) 
  can have the flavor $p=x,y,z$. The $S_{\rm tot}=1$ states are 
  constructed from $S_{\rm tot}=0$ states by replacing one singlet bond 
  with a triplet bond; no nested triplet bond is allowed as explained in the main text.}
\label{fig:spinon_states}
\end{figure}

This can be seen by inspecting the states on a 6-site cluster. 
In Fig.\ \ref{fig:spinon_states} we  represent the $S_{\rm tot}=0$ and 
the \mbox{$S_{\rm tot}=1$} spinon states of a 6-site cluster. 
One can also check that 14 spinon states with $S_{\rm tot}=0$ and 28 ($\times 3$) 
spinon states with $S_{\rm tot}=1$ on an 8-site cluster can be successfully generated
spanning the respective Hilbert subspaces.

\section{Orthonormalization}
\label{sec:ortho}

The spinon basis introduced in the previous section is complete, but not orthonormal. 
In this section, we  orthonormalize the spinon basis so that the basis can be used to 
define second quantized operators in the following. 

We fix the vacuum as it is defined in Eq.\ \reqn{eq:spinon0}. 
To make the 1-spinon states orthonormal to each other we employ the ansatz 
\be 
\label{eq:spinon1_orth_ansatz}
\ket{\Phi_{i}^\sigma} = \alpha_1\ket{\phi_{i}^\sigma} + \alpha_2 \ket{\phi_{i-2}^\sigma}
\ee
and determine the coefficients $\alpha_1$ and $\alpha_2$ such that 
$\langle \Phi_{i}^\sigma | \Phi_{j}^{\sigma'} \rangle=\delta_{i,j}\delta_{\sigma,\sigma'}$
holds. 
The ansatz \reqn{eq:spinon1_orth_ansatz} for the orthonormal 1-spinon states is not unique 
but we did not find a simpler one, i.e., we think there is no other
ansatz involving less sites. Using Eq.\ \reqn{eq:spinon1_ovlp},
we find two solutions 
\bseq
\label{eq:spinon1_o}
\begin{align}
\label{eq:spinon1_r}
\ket{\Phi_{i}^{\sigma,\text{r}}}&=\frac{1}{\sqrt{3}} \left( 2\ket{\phi_{i}^\sigma} + \ket{\phi_{i+2}^\sigma} \right)
\\
\label{eq:spinon1_l}
\ket{\Phi_{i}^{\sigma,\text{l}}}&=\frac{1}{\sqrt{3}} \left( 2\ket{\phi_{i}^\sigma} + \ket{\phi_{i-2}^\sigma} \right)
\end{align}
\eseq
which we call orthonormal ``right'' and ``left'' 1-spinon states, respectively, and denote
by the appropriate superscripts r or l. 

One can choose any of these two solutions to form an orthonormal spinon basis 
\footnote{One can use the ``right'' and the ``left'' 
1-spinon states also combined to construct orthonormal many-spinon states. 
For example, the 2-spinon 
state $\ket{\Phi_{i,i+d}}$ with $d\geq5$ can be constructed from the 
direct product of $\ket{\Phi_{i}^{\sigma,\text{r}}}$ 
and $\ket{\Phi_{i+d}^{\sigma,\text{l}}}$. However, it turns out that this leads to finite overlaps between 
different (for example 2- and 4-) spinon subspaces.}. 
We stress that in the orthonormal 1-spinon state the spinon is not localized 
at a single lattice site, but it has an extension of two lattice spacings to the right, 
$\ket{\Phi_{i}^{\sigma,\text{r}}}$, 
or to the left, $\ket{\Phi_{i}^{\sigma,\text{l}}}$. In the 
following, we use the 1-spinon state ``left'' and drop the index $l$
to lighten the notation: $\ket{\Phi_{i}^{\sigma}}:=\ket{\Phi_{i}^{\sigma,\text{l}}}$. 

The orthonormal many-spinon states with larger distances
between the spinons are constructed from the direct product of orthonormal 1-spinon states.
Since each orthonormal spinon extends over two sites, this construction fails 
if the distance between spinons is 1.
In this case the state is ``distorted'' and we need to perform an explicit orthonormalization to define it
properly. The orthonormal singlet and triplet 2-spinon states for $d\geq 3$ are given by
\begin{subequations}
\begin{alignat}{2}
\label{eq:spinon2_s_orth}
\ket{\Phi_{i,i+d}^{s}}&:=
\sum_{\sigma\sigma'} \chi_{\sigma\sigma'}^{s} \ket{\Phi_{i}^\sigma } \otimes \ket{ \Phi_{i+d}^{\sigma'} }=
\frac{1}{3} \left( 4\ket{\phi_{i,i+d}^{s\vphantom{t,m}}} \right.
 \nn \\ 
&\hspace{-0.5cm} \left. + 2\ket{\phi_{i-2,i+d}^s} + 2\ket{\phi_{i,i+d-2}^s} + 
\ket{\phi_{i-2,i+d-2}^{s\vphantom{t,m}}} \right) ,
\\ 
\label{eq:spinon2_t_orth}
\ket{\Phi_{i,i+d}^{t,p}}&:=
\sum_{\sigma\sigma'} \chi_{\sigma\sigma'}^{t,p} \ket{\Phi_{i}^{\sigma} } \otimes \ket{\Phi_{i+d}^{\sigma'}} 
=\frac{1}{3} \left( 4\ket{\phi_{i,i+d}^{t,p}} \right.
\nn \\
& \hspace{-0.5cm} \left.+ 2\ket{\phi_{i-2,i+d}^{t,p}} + 
2\ket{\phi_{i,i+d-2}^{t,p}} + \ket{\phi_{i-2,i+d-2}^{t,p}} \right),
\end{alignat}
\end{subequations}
where $\chi_{\sigma\sigma'}^{s}$ and $\chi_{\sigma\sigma'}^{t,p}$ are the Clebsch-Gordan coefficients 
for singlet and triplet states. 
The orthonormal triplet state with $d=1$ is such a distorted state. To construct it, we start from 
\be 
\label{eq:spinon2_d1_ans}
\ket{\Phi_{i,i+1}^{t,p}} := \frac{1}{N}\left( \alpha_1 \ket{\phi_{i,i+1}^{t,p}} + \alpha_2 \ket{\phi_{i-2,i+1}^{t,p}} 
+ \ket{\phi_{i-2,i-1}^{t,p}} \right).
\ee
This ansatz is motivated by the extension of each orthonormal spinon by two sites to the left. Requiring
that \reqn{eq:spinon2_d1_ans} is orthonormal to the triplet 2-spinon states leads to the 
non-trivial equations
\begin{subequations}
\begin{align}
\braket{\Phi_{i-d+1,i+1}^{t,p}}{\Phi_{i,i+1}^{t,p}}&=0  \Longrightarrow 2\alpha_1 -3\alpha_2=0 \\ 
\braket{\Phi_{i-d-1,i-1}^{t,p}}{\Phi_{i,i+1}^{t,p}}&=0  \Longrightarrow \phantom{2}\alpha_2 -2=0
\end{align}
\end{subequations}
with $d\geq 3$, and a normalization condition for $N$. One obtains
\be 
\ket{\Phi_{i,i+1}^{t,p}} = \frac{1}{\sqrt{6}}
\left(  
3 \ket{\phi_{i,i+1}^{t,p}} +2 \ket{\phi_{i-2,i+1}^{t,p}} + \ket{\phi_{i-2,i-1}^{t,p}}
\right).
\ee
The orthonormal 3-spinon states are reported in Appendix \ref{ap:spinon3}. 
We notice that the orthonormal $n$-spinon states can always be
constructed from states with the same or a smaller number of spinons.

The many-spinon states constructed from the direct product of orthonormal 1-spinon states  
are not fully orthogonal. Some finite overlaps occur in the  3- and higher spinon subspaces. 
A finite overlap occurs if the positions of orthonormal spinons match. 
For instance, in the 3-spinon sector we find
\be
\label{eq:overlap_spinon3}
\braket{\Phi_{i}^{\sigma}\Phi_{i+n,i+n+d}^{s}}{\Phi_{i,i+n}^{s}\Phi_{i+n+d}^{\sigma}} = 
-\frac{1}{2} \quad; \quad n\geq 3.
\ee
There is no such overlap in the 1-spinon and 2-spinon subspaces
 because after fixing the positions of the spinons there is only 
one possibilty to form states with specific total spin and flavor.
But in the 3-spinon subspace there are two possibilities, 
see the bra and the ket states in \reqn{eq:overlap_spinon3}. 
This finite overlap is not a serious issue because subspaces with different numbers of spinons
 are mutually orthogonal. 
But to define the proper interactions in second quantization one has to
 carefully take the finite overlaps into account, see Appendix \ref{ap:sec:int_corff}.

All spinon states can be expanded in orthonormal states. For the 1-spinon state \reqn{eq:spinon1} and  
the singlet 2-spinon state \reqn{eq:spinon2_s} we find
\bseq 
\label{eq:expanI}
\begin{align}
\label{eq:expanI:a}
\ket{\phi_{i}^{\sigma} }
&=\frac{\sqrt{3}}{2} \sum_{m\geq 0} 
\left(-\frac{1}{2} \right)^{\frac{m}{2}} \ket{\Phi_{i-m}^{\sigma}}, 
\\
\label{eq:expanI:b}
\ket{\phi_{i,i+d}^{s} } 
&=-\frac{1}{2}\ket{\phi_{i,i+d-2}^{s}}+\frac{3}{4} \sum_{m\geq 0} 
\left(-\frac{1}{2} \right)^{\frac{m}{2}} \ket{\Phi_{i-m,i+d}^s}. 
\end{align}
\eseq
with $m$ even. These relations are obtained by reversing Eqs. \reqn{eq:spinon1_l} and \reqn{eq:spinon2_s_orth}. 
Eq. \reqn{eq:expanI:b} is to be used in a recursive way starting from $d=3$. We notice that $\ket{\phi_{i,i+1}^{s}}:=\ket{0}$.
A local spinon state represented in terms of orthonormal states becomes extended over the 
whole chain with a prefactor decreasing exponentially for increasing distance. 
Generally, in the expansion of a $m$-spinon state orthonormal states appear with $m$ or less spinons.

\section{Spinon-pair operators}
\label{sec:pair-operators}

\subsection{Definition}

Any attempt to define single spinon creation or annihilation operators 
runs into severe problems. Typically, one has to work in a larger Hilbert
space, for instance enlarge it artificially, and complement the
description by a severe constraint so that the accessible Hilbert space
is again the physical one \cite{wen91}.

In the present article, we want to follow a different route. We refrain from
defining a single creation or annihilation event, but define spinon operators 
for pairs of spinons in a rather straightforward manner. No severe constraints
are required to deal with the physical Hilbert space because creation and annihilation 
of spinons always happen in pairs. Spinon hopping does not alter the number of spinons
so that it can also be expressed by a second quantized operator which
addresses the hopping process as a whole. 

There is an important point that has to be clarified 
before defining the spinon-pair operators. 
Considering a segment of the chain, there can be two local vacua $\ket{0_{i,i+d}}$ and 
$\ket{\tilde{0}_{i,i+d}}$ given by 
\bseq 
\begin{align}
\label{eq:right_vac}
 \ket{0_{i,i+d}^\pdag}&:=
 \includegraphics[trim= 0.5 1.4 0 0.4, width=0.25\textwidth]{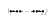}
 , \\
\label{eq:wrong_vac}
 \ket{\tilde{0}_{i,i+d}^\pdag}_{\vphantom{\dagger}}
 &:=
 \includegraphics[trim= 0.8 1.44 0 -0.5, width=0.25\textwidth]{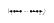}.
\end{align}
\eseq
Creation of 
two spinons at positions $i$ and $i+d$ from $\ket{0_{i,i+d}}$ changes 
the state on the chain between sites $i$ and $i+d$. However, a two-spinon creation from 
$\ket{\tilde{0}_{i,i+d}}$ corresponds to a change in the state of the chain \emph{before }
site $i$ and \emph{beyond} site $i+d$. We call $\ket{0_{i,i+d}}$ ``right'' local vacuum and
 $\ket{\tilde{0}_{i,i+d}}$ ``wrong'' local vacuum.

The singlet operator $\mathcal{S}_{i,i+d}^{\dagger}$ with $d\geq3$ is defined 
to create two orthonormal spinons 
at sites $i$ and $i+d$ with total spin zero 
if the state on the chain between sites $i$ and $i+d$ 
is the right local vacuum $\ket{0_{i,i+d}}$. This operator can be expressed as
\be
\mathcal{S}_{i,i+d}^{\dagger}: \ket{0_{i,i+d}^\vpdag}
\longrightarrow 
\includegraphics[trim= 0 1.4 0 0, width=0.22\textwidth]{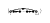}
\label{eq:op_sing}
\ee 
where the states before $i$ and beyond $i+d$ are supposed to be arbitrary;
they are not changed.
The orthonormal spinons at positions $i$ and $i+d$ are indicated by empty circles.
The result of $\mathcal{S}_{i,i+d}^{\dagger}$ is zero if there is \emph{any} other 
state different than $\ket{0_{i,i+d}^\vpdag}$ \emph{between} sites $i$ and $i+d$. Hence, the
singlet operator is defined only over odd distances $d$. 
We define $\mathcal{S}_{i,i+1}^{\dagger}:=0$.

Depending on the state on the chain sites $i-2$, $i-1$, $i+1$, and $i+2$, namely
if they  are occupied by spinons or not, the final state is distorted or not,
as defined above. It is apparent from this definition that the singlet operator 
$\mathcal{S}_{i,i+d}^{(\dagger)}$ is a string operator of which the  action 
depends on the state between sites $i-2$ to $i+d+2$. Moreover, we notice 
that there is no need to introduce additional operators to create distorted states 
at small distances. Each state from the $S_{\rm tot}=0$ ($S_{\rm tot}=\frac{1}{2}$) 
subspace can be generated by applying the singlet operators on the spinon vacuum (a 1-spinon state).

The order of singlet operators does not matter in the 
creation of distinct singlets. However, to create nested singlets one has to start 
from the outermost singlet. This implies that two creation (annihilation) 
singlet operators do not necessarily commute.

We define the triplet operator $\mathcal{T}_{i,i+d}^{p\, \dagger}$ with $d\geq1$ 
which creates a triplet bond with flavor $p$ between sites $i$ and $i+d$ from the 
right local vacuum $\ket{0_{i,i+d}^\vpdag}$. Similar to the singlet operator, 
the result is zero if there is any state different than $\ket{0_{i,i+d}^\vpdag}$ 
between $i$ and $i+d$. \emph{In addition}, the result is zero if the action of the 
triplet operator leads to the creation of a nested triplet bond. 
This means that the application of the triplet operator 
depends on the state on the whole chain and not only on the state between sites $i$ and $i+d$.
This global feature of the triplet operator limits its general applicability. However, one 
can approximate it by its leading local contribution. The term ``leading''
refers to an expansion in terms with non-zero action on an increasing number of spinons.

First, we notice that $\mathcal{T}_{i,i+d}^{p\,\dagger}$ with 
$i$ odd (or even, depending on how we label the chain sites) 
always leads to creation of nested triplet bonds and hence it is zero. 
The action of $\mathcal{T}_{i,i+d}^{p\,\dagger}$ with $i$ even can still lead 
to nested triplet bonds, {\it but} this requires at least four spinons 
(two nested singlets) in the system. Hence the leading contribution of the triplet operator can be described by the local action 
\be
\mathcal{T}_{i,i+d}^{p\,\dagger}: \ket{0_{i,i+d}^\vpdag} 
\longrightarrow 
\includegraphics[trim= 0 1.07 0 0, width=0.21\textwidth]{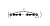}
\label{eq:op_trip}
\ee
for $i$ even and zero for $i$ odd. 
In this argument, we suppose arbitrary states on the chain before site $i$ 
and beyond site $i+d$. The whole $S_{\rm tot}=1$ subspace can be generated 
by applying the singlet \reqn{eq:op_sing} and the triplet \reqn{eq:op_trip} operators 
to the vacuum. There will be some redundant states, i.e.,
some overcounting occurs. But this will only happen where at least  6 spinons 
are present  due to the approximation \reqn{eq:op_trip}. 
Hence we accept this degree of overcounting 
because it matters only on the hexatic level of operators. 
At low densities of spinons it is irrelevant.

The hopping operator $\mathcal{H}_{i,i+d}^{\sigma\sigma'}$ annihilates an orthonormal spinon 
with spin $\sigma$ at position $i$ and creates an orthonormal spinon with 
spin $\sigma'$ at position $i+d$ if there is the right 
local vacuum between $i+1$ and $i+d$; otherwise  the result is zero.
This restricts the hopping distance $d$ to even values. 
One notices that the crossing of spinons is prohibited, i.e., 
the result is zero if there is any spinon between sites $i$ and $i+d$.  
We obtain
\be
\mathcal{H}_{i,i+d}^{\uparrow\uparrow}: 
\includegraphics[trim= 1. 1.2 1.1 0.3, width=0.156\textwidth]{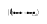} 
\longrightarrow 
\includegraphics[trim= 1. 1.2 1.1 0.3, width=0.156\textwidth]{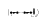} 
\label{eq:hop_op}
\ee 
and similarly for other values $\sigma$ and $\sigma'$.

Instead of introducing hopping operators $\mathcal{H}_{i,i+d}^{\sigma\sigma'}$ 
with specific spin indices $\sigma$ and $\sigma'$ it turns out to be more convenient to
define
\bseq
\begin{align}
\label{eq:hop_op_n}
\mathcal{H}_{j,j+d}^\pdag &:= \mathcal{H}_{j,j+d}^{\uparrow\uparrow}+
\mathcal{H}_{j,j+d}^{\downarrow\downarrow}, \\ 
\label{eq:hop_op_x}
\mathcal{H}_{j,j+d}^x &:= \mathcal{H}_{j,j+d}^{\uparrow\downarrow}+
\mathcal{H}_{j,j+d}^{\downarrow\uparrow}, 
\\
\label{eq:hop_op_y}
\mathcal{H}_{j,j+d}^y &:= i\left(\mathcal{H}_{j,j+d}^{\uparrow\downarrow}-
\mathcal{H}_{j,j+d}^{\downarrow\uparrow} \right), \\
\label{eq:hop_op_z}
\mathcal{H}_{j,j+d}^z &:= \mathcal{H}_{j,j+d}^{\uparrow\uparrow}-
\mathcal{H}_{j,j+d}^{\downarrow\downarrow}. 
\end{align}
\eseq
Note that in a SU(2) invariant model no hopping with spin flips 
such as $\mathcal{H}_{j,j+d}^{\downarrow\uparrow}$ will occur.
But even in such model,  the hopping operators with
spin flips appear in products of intermediate calculations, see below.

The action of the neutral hopping operator \reqn{eq:hop_op_n} 
on a singlet (triplet) bond is always a singlet (triplet) bond.  
The flavor hopping operators $\mathcal{H}_{j,j+d}^p$ with $p=x,y,z$ 
acting on a singlet bond change it into a triplet bond with flavor $p$ and vice versa. 
We note that the action of the neutral hopping operator $\mathcal{H}_{j,j+d}^\pdag$ 
only depends on the state between sites $j$ and $j+d$. The actions of 
the flavor hopping operators depend on the state on the whole chains
 because they could  replace a nested 
singlet bond with a triplet bond which is not allowed.

We call the singlet operators, the triplet operators, and the hopping operators 
bilinear although they are basically string operators. We do so because by analogy to
a fermionic or bosonic Hamiltonian. The key property is that they 
are characterized by their action at two sites on the chain.

The quartic interactions are given by the normal-ordered product of 
two bilinear operators. To describe spinon interactions up to the quartic level in a SU$(2)$-symmetric Hamiltonian only the singlet operators, the triplet operators, and the 
neutral hopping operators are required.  
The polarized hopping operators can only appear on higher levels of interactions or 
in the intermediate steps of the calculations, for instance, as the results of commutators.

\subsection{Algebra}

The next point to address is the commutation of different spinon-pair operators. 
Explicitly, we are interested in the normal-ordered form of 
$\left[ \mathcal{S}^{\dagger}, \mathcal{H} \right]$, 
$\left[ \mathcal{S}^{\dagger}, \mathcal{S} \right]$, 
$\left[ \mathcal{S}^{\dagger}, \mathcal{S}^{\dagger} \right]$, 
$\left[ \mathcal{S}^{\dagger}, \mathcal{T} \right]$, and 
$\left[ \mathcal{H}, \mathcal{H} \right]$. By normal-ordering we mean 
that we sort the effect of the commutators according to the number
of spinons needed for the term to become active, i.e., to have
a non-trivial effect. The 
commutators are calculated by inspecting their effects on arbitrary states.

\begin{figure}[thb]
  \centering
  \includegraphics[trim=0 0.2 0 0.5, width=0.97\columnwidth,angle=0]{fig16a}
  \caption{(Color online) Analysis of the commutator 
 $[ \mathcal{S}_{i,i+d}^{\dagger}, \mathcal{H}_{j,j+d'}^{\protect\phantom{\dagger}} ]$ 
in different cases. The double-headed arrow creates two spinons at the two ends 
($i$ and $i+d$) and denotes the singlet operator $\mathcal{S}_{i,i+d}^{\dagger}$. 
The hopping operator  $\mathcal{H}_{j,j+d'}^{\protect\pdag}$ is shown by 
a dashed arrow indicating that a spinon hops from site $j$ to $j+d'$.}
\label{fig:comm_SHp}
\end{figure}

Let us start with the commutator 
$[ \mathcal{S}_{i,i+d}^{\dagger}, \mathcal{H}_{j,j+d'}^\vpdag ]$. 
We restrict $d'$ to positive values; negative values will be discussed below. 
Different cases are schematically distinguished in Fig.\ \ref{fig:comm_SHp}. 
The singlet operator $\mathcal{S}_{i,i+d}^{\dagger}$ in this figure is represented 
by a double-headed arrow;  two spinons are created at the two tips, $i$ and $i+d$. 
The hopping operator $\mathcal{H}_{j,j+d'}^\vpdag$ is depicted by a dashed arrow 
which specifies the hopping process from $j$ to $j+d'$. 

If $j<i$ and $j+d' \leq i+d$ or if $j>i+d$ the result will be zero considering the 
properties of the singlet operators and the hopping operators. 
The result will be also zero if $i<j,j+d'\leq i+d$. 
If $j<i+d$ and $j+d'>i+d$, the second part of the commutator vanishes 
and we obtain  
$[ \mathcal{S}_{i,i+d}^{\dagger}, \mathcal{H}_{j,j+d'}^\vpdag ]=
\mathcal{S}_{i,i+d}^{\dagger}\mathcal{H}_{j,j+d'}^\vpdag$.
For $j=i+d$, the first part vanishes
$\mathcal{S}_{i,i+d}^{\dagger}\mathcal{H}_{j,j+d'}^\vpdag=0$ and 
the commutator equals $-\mathcal{S}_{i,i+d+d'}^{\dagger}$.
Finally, if $j=i$ and $j+d'<i+d$, the commutator simplifies to 
$-\mathcal{P}_{i,i+d'-1}^\vpdag\mathcal{S}_{i+d',i+d}^{\dagger}$.
 
The projection operator $\mathcal{P}_{i,i+d}^\vpdag$ is defined to be
identity if there is the right local vacuum \reqn{eq:right_vac} 
between sites $i$ and $i+d$ and zero otherwise. 
We define $\mathcal{P}_{i,i-1}^\vpdag:=1$. 
Combining everything, we obtain
\begin{align}
\label{eq:comm_SHp}
\left[ \!\!\vphantom{\mathcal{S}_{i,i+d}^{\dagger}} \right.&\left. 
\mathcal{S}_{i,i+d}^{\dagger}, 
\mathcal{H}_{j,j+d'}^\vpdag \right] 
= -\mathcal{S}_{i,i+d+d'}^{\dagger} \delta_{i+d,j}^\vpdag \nn \\
&-\mathcal{P}_{i,i+d'-1}^\vpdag\mathcal{S}_{i+d',i+d}^{\dagger} 
\theta(d-d')\delta_{i,j}^\vpdag  \nn 
\\
&+\mathcal{S}_{i,i+d}^{\dagger}\mathcal{H}_{j,j+d'}^\vpdag  
\theta(i+d-j)  \theta(j+d'-i-d),
\end{align}
where the step function $\theta(x)$ is 1 for $x>0$ and zero for $x \leq 0$. 

It is instructive to compare Eq.\ \reqn{eq:comm_SHp} 
to its counterpart for hardcore bosons. For the hardcore boson $b_i^{(\dagger)}$ 
acting on site $i$ one has
\begin{align}
\label{eq:comm_SHp_b}
\left[ b_{i}^{\dagger}b_{i+d}^{\dagger}, b_{j+d'}^{\dagger} b_{j}^\vpdag \right]
=&-b_{i}^{\dagger}b_{i+d+d'}^{\dagger}(1-b_{i+d}^{\dagger}b_{i+d}^\vpdag) 
\delta_{i+d,j}^\vpdag \nn \\
&-(1-b_{i}^{\dagger}b_{i}^\vpdag)b_{i+d'}^{\dagger}b_{i+d}^{\dagger} 
\delta_{i,j}^\vpdag \nn \\
&+ b_{i}^{\dagger}b_{i+d}^{\dagger}b_{j+d'}^{\dagger}b_{j}^\vpdag 
(\delta_{i+d,j}^\vpdag+\delta_{i,j}^\vpdag),
\end{align}
where the local projection operator $(1-b_{i}^{\dagger}b_{i}^\vpdag)$ guarantees that site $i$ is empty. 
We compare the right hand side of the two Eqs.\ \reqn{eq:comm_SHp} and \reqn{eq:comm_SHp_b} term by term. The first term in Eq.\ \reqn{eq:comm_SHp} 
is similar to the one of \reqn{eq:comm_SHp_b} except that the projection operator is 
absorbed in the definition of the singlet operator 
because it occurs between sites $i$ and $i+d+d'$. 
The local projection operator $(1-b_{i}^{\dagger}b_{i}^\vpdag)$ 
in the second term of \reqn{eq:comm_SHp_b} is replaced by $\mathcal{P}_{i,i+d'-1}^\vpdag$ in \reqn{eq:comm_SHp}. 
In addition, the step function $\theta(d-d')$ in the second term of \reqn{eq:comm_SHp} 
reflects the fact that spinons cannot cross each other while bosons can. The step 
functions in the third term of \reqn{eq:comm_SHp} instead of the local delta functions 
in \reqn{eq:comm_SHp_b} also stem from the fact that spinons cannot pass each other
on the chain.

Similarly, the commutator $[ \mathcal{S}_{i,i+d}^{\dagger}, \mathcal{H}_{j+d',j}^\vpdag ]$
is analyzed with $d' \geq0$ leading to 
\begin{align}
\label{eq:comm_SHn}
\left[ \mathcal{S}_{i,i+d}^{\dagger},\right. \!\!&\left. \mathcal{H}_{j+d',j}^\vpdag \right] 
= -\mathcal{S}_{i-d',i+d}^{\dagger} \delta_{i,j+d'}^\vpdag \nn \\
&-\mathcal{S}_{i,i+d-d'}^{\dagger}\mathcal{P}_{i+d-d'+1,i+d}^\vpdag \theta(d-d')
\delta_{i+d,j+d'}^\vpdag  \nn \\
&+\mathcal{S}_{i,i+d}^{\dagger}\mathcal{H}_{j+d',j}^\vpdag  \theta(i-j) \theta(j+d'-i).
\end{align}
The commutator of creation and annihilation singlet operators is given by
\begin{align}
\label{eq:comm_SdS}
\left[ \mathcal{S}_{i,i+d}^{\dagger}, \right. \!\!& \left. \mathcal{S}_{j,j+d'}^\vpdag \right] 
= - \mathcal{P}_{i,i+d}^\vpdag \delta_{i,j}^\vpdag\delta_{d,d'}^\vpdag \nn \\
&+\frac{1}{2} 
\left( 
\mathcal{H}_{i+d+d',i}^\vpdag \delta_{j,i+d}^\vpdag + 
\mathcal{H}_{i-d',i+d}^\vpdag \delta_{j,i-d'}^\vpdag \right)
\nn \\
&+ \mathcal{S}_{i,i+d}^{\dagger} \mathcal{S}_{j,j+d'}^\vpdag \theta(i+d-j)  \theta(j+d'-i).
\end{align}
The leading contribution of this commutator is the projection operator 
$\mathcal{P}_{i,i+d}^\vpdag$ which would reduce to the identity if the operators 
were normal fermions or bosons. The prefactor $\nicefrac{1}{2}$ in the second 
line results from the normalization of singlet states.

One can also check that the commutator between the creation singlet operator and the annihilation triplet operator reads
\begin{align}
\label{eq:comm_SdT}
 \left[ \mathcal{S}^{\dagger}_{i,i+d}, \right. \!\!& \left. \mathcal{T}^{p}_{j,j+d'} \right]
= \frac{1}{2} 
\left(
\delta_{i,j+d'}^{\vphantom{\dagger}} \mathcal{H}_{i-d',i+d}^{p} 
-\delta_{j,i+d}^{\vphantom{\dagger}} \mathcal{H}_{i+d+d',i}^{p} 
\right)  
\nn \\ &
+ \mathcal{S}^{\dagger}_{i,i+d} \mathcal{T}^{p}_{j,j+d'} \theta(i+d-j)\theta(j+d'-i).
\end{align}
Other useful commutators are provided in App.\ \ref{ap:comm}. 

\subsection{Projection Operator}

In practical calculation, the projection operator has to be expressed 
in terms of the singlet operators, the triplet operators, and the hopping operators. 
The projection operator $\mathcal{P}_{\!i,i+d}^\vpdag$ is zero 
if there is a spinon at or between sites $i$ and $i+d$. In addition, the
projection returns zero if there is the wrong local vacuum \reqn{eq:wrong_vac} 
between $i$ and $i+d$.
The former property can be simply captured by 
\be 
\prod_{j=i}^{i+d}(1-\mac{H}_{j,j}^\pdag)
\ee
which vanishes if there exists a spinon at or between $i$ and $i+d$.
The latter property, however, which requires distinguishing between 
the right \reqn{eq:right_vac} and the wrong \reqn{eq:wrong_vac} local vacua, is not 
a local feature. This makes it difficult to find a representation for the 
projection operator.

Nevertheless, the projection operator never appears alone in our calculations. It 
either occurs in a sum over chain sites or it is multiplied with other operators, 
see the following. In these cases, one can find the leading contributions of 
the expression by applying it to the first subspaces containing only a few spinons.

We start with the operator 
$\mac{P}_{\!d}^{\vphantom{\dagger}}:=\sum_i\mathcal{P}_{i,i+d}^\vpdag$ 
where the sum $i$ runs over the chain sites. To find the leading contributions 
of $\mac{P}_{\!d}^{\vphantom{\dagger}}$ we consider the ansatz 
\begin{align}
\label{eq:Pd_ans}
\mathcal{P}_{\!d}^{\vphantom{\dagger}} = C_0^{\vphantom{\dagger}} \sum_i \mathds{1}  
\!&+C_1^{\vphantom{\dagger}} \sum_i \mathcal{H}_{i,i}^{\vphantom{\dagger}}
\!+ \!\sum_i \sum_{d'} C_2^{\vphantom{\dagger}}(d') \!
\left( 
\mathcal{S}_{i,i+d'}^{\dagger} \mathcal{S}_{i,i+d'}^\vpdag 
\right. \nn \\ & 
+\!\!\!\sum_{p=x,y,z}\!\!\mathcal{T}_{i,i+d'}^{p\,\dagger} \mathcal{T}_{i,i+d'}^{p} 
\left. \vphantom{\mathcal{S}_{i,i+d'}^{\dagger}} \right) + \cdots
\end{align}
where ``$\cdots$'' stands for $3-$ and higher spinon interaction terms.
The prefactor $C_0^{\vphantom{\dagger}}$ is determined by applying the 
relation \reqn{eq:Pd_ans} to the spinon vacuum. 
We obtain $\bra{0}\mathcal{P}_{d}^{\vphantom{\dagger}} \ket{0} 
=\frac{L}{2}$ which yields $C_0^{\vphantom{\dagger}}=\frac{1}{2}$.
To calculate the prefactor $C_1^{\vphantom{\dagger}}$ one needs to apply 
$\mac{P}_{\!\!d}^{\vphantom{\dagger}}$ to the 1-spinon state. 
We have $\bra{\Phi_{i}^{\sigma}} \mathcal{P}_{\!d}^{\vphantom{\dagger}} 
\ket{\Phi_{i}^{\sigma}}=(\frac{L}{2}-\frac{d}{2})$ 
which leads to $C_1^{\vphantom{\dagger}}=-\frac{d}{2}$. 
To find the interaction potential $C_2^{\vphantom{\dagger}}(d')$, 
one needs to compute $\bra{\Phi_{i,i+d'}^{s}} \mathcal{P}_{d}^{\vphantom{\dagger}} 
\ket{\Phi_{i,i+d'}^{s}}$.
We identify $C_2^{\vphantom{\dagger}}(d')=\frac{d-d'}{2}$ for $d' \leq d$ and zero otherwise. Therefore, the final result reads
\begin{align}
\label{eq:Pd_exp}
\mathcal{P}_{\!d}^{\vphantom{\dagger}} = \frac{1}{2} \sum_i \mathds{1}  
\!&-\frac{d}{2} \sum_i \mathcal{H}_{i,i}^{\vphantom{\dagger}}
\!+ \!\sum_i \sum_{d'\leq d} \frac{d-d'}{2} \!
\left( 
\mathcal{S}_{i,i+d'}^{\dagger} \mathcal{S}_{i,i+d'}^\vpdag 
\right. \nn \\ & 
+\!\!\!\sum_{p=x,y,z}\!\!\mathcal{T}_{i,i+d'}^{p\,\dagger} \mathcal{T}_{i,i+d'}^{p} 
\left. \vphantom{\mathcal{S}_{i,i+d'}^{\dagger}} \right) + \cdots \quad.
\end{align}
It is remarkable that the prefactor of the spinon density operator 
$\mathcal{H}_{i,i}^{\vphantom{\dagger}}$ is proportional to the distance $d$. 
The two-spinon interaction potential in $\mathcal{P}_{\!d}$  decreases linearly
with increasing distance between the spinons and vanishes at the maximum distance $d$. 

The product of the projection operator and the singlet operator also appears in the commutators, see for example Eqs.\  \reqn{eq:comm_SHp} and \reqn{eq:comm_SHn}. We consider 
$\mathcal{S}_{i+d_\text{e},i+d_\text{e}+d'}^{\dagger} \mathcal{P}_{\!i,i+d}^\vpdag$. 
Here and in the following we use the subscripts ``e'' and ``o'' to indicate even and odd
 numbers. We observe that
\be
\mathcal{S}_{i+d_\text{e},i+d_\text{e}+d'}^{\dagger} \mathcal{P}_{\!i,i+d}^\vpdag \ket{0}
=\mathcal{S}_{i+d_\text{e},i+d_\text{e}+d'}^{\dagger} \ket{0}
\ee
because the singlet operator guarantees the existence of the 
right local vacuum between $i$ and $i+d$.
To study the effect of the product on 1-spinon states we distinguish three cases: 
(i) $d_\text{e}\geq 0$ and $d_\text{e}+d'\geq d$ 
(ii) $d_\text{e}\geq 0$ and 
$d_\text{e}+d'< d$ (iii) $d_\text{e}<0$. 
We analyze case (i) explicitly; the other cases can be treated in the same way.

The action of $\mathcal{S}_{i+d_\text{e},i+d_\text{e}+d'}^{\dagger} 
\mathcal{P}_{\!i,i+d}^\vpdag$ 
on the 1-spinon state $\ket{\Phi_{j}^{\sigma}}$ is given by 
\begin{align} 
\mathcal{S}_{i+d_\text{e},i+d_\text{e}+d'}^{\dagger} \mathcal{P}_{\!i,i+d}^\vpdag 
\ket{\Phi_{j}^{\sigma}}
=&\left( 1-\theta(j-i+1)\theta(i+d_\text{e}-j)^{\vphantom{\dagger}} \right) 
\nn \\ &
\times \mathcal{S}_{i+d_\text{e},i+d_\text{e}+d'}^{\dagger} \ket{\Phi_{j}^{\sigma}}.
\end{align}
We notice that even for $i+d<j<i+d_\text{e}$ the result is zero because for $(j-i)$ odd 
the wrong local vacuum appears between $i$ and $i+d$ and for $(j-i)$ even 
the wrong local vacuum appears between $i+d_\text{e}$ and $i+d_\text{e}+d'$.  
Therefore, we obtain
\begin{align}
\label{eq:SdP_exp_i}
\!\!\!\mathcal{S}_{i+d_\text{e},i+d_\text{e}+d'}^{\dagger} &
\mathcal{P}_{\!i,i+d}^{\vphantom{\dagger}} =
 \mathcal{S}_{i+d_\text{e},i+d_\text{e}+d'}^{\dagger} \left( 1  
-\!\!\sum_{n_\text{o}=1}^{d_\text{e}-1} 
\mathcal{H}_{i+n_\text{o}^\vpdag,i+n_\text{o}^\vpdag}^{\vphantom{\dagger}} 
\right. \nn \\ & 
\left. \vphantom{\sum_{m_\text{o}=1}^{n-1}}
+ \cdots \right)
\, ; \, \{ d_\text{e}\geq 0 ~{\rm and}~ d_\text{e}+d' \geq d \},
\end{align}
where ``$\cdots$'' involves 2- and higher spinon interactions which we  neglect. 
The sum over $n_\text{o}$ is limited to odd numbers because for even numbers  
the wrong local vacuum occurs between $i+d_\text{e}$ and 
$i+d_\text{e}+d'$ and the vanishing is guaranteed by the application of the 
singlet operator $\mathcal{S}_{i+d_\text{e},i+d_\text{e}+d'}^{\dagger}$.  

One can analyze cases (ii) and (iii) to obtain  relations analogous to
\reqn{eq:SdP_exp_i}. Combining the three equations  
and after some simplifications we obtain 
\begin{align}
\label{eq:SdP_exp}
&\mathcal{S}_{i+d_\text{e},i+d_\text{e}+d'}^{\dagger} 
\mathcal{P}_{\!i,i+d}^{\vphantom{\dagger}} =
 \mathcal{S}_{i+d_\text{e},i+d_\text{e}+d'}^{\dagger} \left( 1  \vphantom{\sum}
 \right. \nn \\ & 
 \quad \quad \quad \quad
-\sum\limits_{n_\text{e}} \mathcal{H}_{i+n_\text{e}^\pdag,i+n_\text{e}^\pdag}^\vpdag  
\theta(d\!-\!n_\text{e})\theta(n_\text{e}\!-\!d_\text{e}\!-\!d') 
\nn \\ &
 \quad \quad \quad \quad
-\sum_{n_\text{o}} \mathcal{H}_{i+n_\text{o}^\pdag,i+n_\text{o}^\pdag}^\vpdag 
\theta(d_\text{e}\!-\!n_\text{o}) \theta(n_\text{o})
\left. \vphantom{\sum}
+ \cdots \right)
\end{align}
which is valid up to quartic level in spinon creation and annihilation operators.

We also inspect the product of the neutral hopping operator with 
the projection operator: $\mathcal{P}_{i+d_\text{e}+n,i+d_\text{e}+n+d}^{\vpdag} 
\mathcal{H}_{i,i+d_\text{e}}^{\vphantom{\dagger}}$ where $n$ is 
odd for $n>0$ and even for $n<0$. The expression can be analyzed by 
applying it to the 1- and 2-spinon sectors similar to the above discussion. 
The final result reads
\allowdisplaybreaks
\begin{align}
\label{eq:PH_exp}
&\mathcal{P}_{i+d_\text{e}+n,i+d_\text{e}+n+d}^{\vpdag} 
\mathcal{H}_{i,i+d_\text{e}}^{\vphantom{\dagger}} 
= \theta(n) \!\left( \vphantom{\sum_m} \mathcal{H}_{i,i+d_\text{e}}^\vpdag 
\right. \nn \\ &
-\!\!\sum\limits_{m_\text{o}\geq 1}\!\!\theta(m_\text{o}\!+\!d_\text{e})
\theta(n\!+\!d\!-\!m_\text{o})
\!\left(
\mathcal{S}_{i+d_\text{e},i+d_\text{e}+m_\text{o}}^{\dagger} 
\mathcal{S}_{i,i+d_\text{e}+m_\text{o}}^\vpdag 
\right. \nn \\ & 
+\!\!\!\!\sum_{p=x,y,z}\!\!\mathcal{T}_{i+d_\text{e},i+d_\text{e}+m_\text{o}}^{p\,\dagger} 
\mathcal{T}_{i,i+d_\text{e}+m_\text{o}}^{p} 
\left. \vphantom{\mathcal{S}_{i,i+d'}^{\dagger}} \right) 
\!\!\left. \vphantom{\sum_m} \right)
+\theta(\!-n\!-\!d) \!\left( \vphantom{\sum_m} \mathcal{H}_{i,i+d_\text{e}}^\vpdag 
\right. \nn \\ &
-\!\!\sum\limits_{m_\text{o}\geq 1}\!\!\theta(m_\text{o}\!+\!d_\text{e})
\theta(-n\!-\!d_\text{e}\!-\!m_\text{o})
\!\left(
\mathcal{S}_{i-m_\text{o},i+d_\text{e}}^{\dagger} \mathcal{S}_{i-m_\text{o},i}^\vpdag 
\right. \nn \\ & 
+\!\!\!\sum_{p=x,y,z}\!\!\mathcal{T}_{i-m_\text{o},i+d_\text{e}}^{p\,\dagger} 
\mathcal{T}_{i-m_\text{o},i}^{p} 
\left. \vphantom{\mathcal{S}_{i,i+d'}^{\dagger}} \right) 
\left. \vphantom{\sum_m} \right)
+\cdots
\end{align}
where 3- and higher spinon interaction terms are ignored.

Before closing this section, we state that the product of two neutral hopping operators 
is not a valid representation for 2-spinon interactions. In fact, 2-spinon interactions 
are always described in terms of singlet and triplet operators. 
Wherever  the product of two neutral hopping operators appears in the calculations
it has be expanded in terms of hopping operators, singlet operators and triplet operators 
as we did above in the expansion of the projection 
operator. Further useful expansions can be found in App.\ \ref{ap:HH}.

\section{Frustrated Heisenberg chain}
\label{sec:chain}

As an example of the application of the spinon-pair operator representation we study the frustrated Heisenberg chain. Its Hamiltonian has been given in Eq.\ \eqref{eq:j1j2}.
For a chain with an even number of sites, the exact ground state at 
$\alpha=\frac{1}{2}$, known as Majumar-Ghosh point, is given by the 
fully dimerized state \reqn{eq:spinon0} \cite{Majumdar1969a, Majumdar1969b}. 
Upon decreasing the degree of frustration below $\alpha=\frac{1}{2}$ 
the spontaneously dimerized phase undergoes a  second order phase transition to 
a Mott insulator with quasi long-range magnetic order (spin liquid)
at $\alpha_c = 0.241167$ \cite{Okamoto1992,Eggert1996}. 
In both, the  dimerized and the spin liquid phase, the elementary excitations 
(quasiparticles) are known to be spinons \cite{Faddeev1981,Shastry1981}. 
The minimum of the spinon dispersion for chains
with an odd number of sites moves from commensurate to incommensurate momenta 
beyond the Lifshitz point $\alpha_l = 0.538(1)$ \cite{Deschner2013}. 
For systems with an even number of sites, the Lifshitz point takes place 
at smaller frustration $\alpha_l = 0.52036(6)$ \cite{Bursill1995}.

To express the Hamiltonian \reqn{eq:j1j2} in terms of spinon-pair operators, 
we start from the following general ansatz
\begin{align}
\label{eq:j1j2_srep}
 H= &\mathcal{A}_{0:0}^\vpdag H_{0:0}^\vpdag 
+ \mathcal{A}_{1:1}^\vpdag(0) H_{1:1}^\vpdag(0)
+ \! \sum_{d_\text{e}\geq 2} \! \mathcal{A}_{1:1}^\vpdag(d_{e}) 
H_{1:1}^\vpdag(d_\text{e})
\nn \\ &
+ \! \sum_{d_{\rm o}\geq 3}  \mathcal{A}_{2:0}^\pdag(d_\text{o}) 
( H_{2:0}^\vpdag(d_\text{o}) +{\rm h.c.} )
\nn \\ &
+ \! \sum_{d_{o} \geq 3} \sum_{d_{e}} \sum_n 
\mathcal{A}_{3:1}^\vpdag (d_{e},n,d_{o}) 
  ( H_{3:1}^\vpdag(d_{e},n,d_{o}) +{\rm h.c.} )
\nn \\ &
+ \!\! \sum_{d_{o} \geq 3} \sum_{d'_{o} \geq 3} \sum_{d_\text{e}} 
\mathcal{A}_{2:2}^s(d_{\rm o},d_\text{e},d'_{\rm o})
  H_{2:2}^s(d_{\rm o},d_\text{e},d'_{\rm o})
\nn \\ &
+ \!\! \sum_{d_{o} \geq 3} \sum_{d'_{o} \geq 3} \sum_{d_\text{e}} 
\mathcal{A}_{2:2}^t(d_{\rm o},d_\text{e},d'_{\rm o})
  H_{2:2}^t(d_{\rm o},d_\text{e},d'_{\rm o}) 
\nn \\ &
+\cdots
\end{align}
where we defined 
\bseq
\label{eq:srep}
\begin{align}
\label{eq:srep:a}
 H_{0:0}^\vpdag&:=\sum_i \mathds{1}
\\ 
\label{eq:srep:b}
H_{1:1}^\vpdag(0)&:=\sum_i \mathcal{H}_{i,i}^\pdag
 \\ 
\label{eq:srep:c}
H_{1:1}^\vpdag(d_\text{e})&:=\sum_i \mathcal{H}_{i,i+d_{e}}^\pdag+{\rm h.c.} \quad ; 
\quad d_\text{e}\geq2
\\
\label{eq:srep:d}
H_{2:0}^\vpdag(d_\text{o}) &:=\sum_i \mathcal{S}_{i,i+d_{\rm o}}^\dagger
\\
\label{eq:srep:e}
H_{3:1}^\vpdag(d_{e},n,d_{o})&:=\sum_i\mathcal{S}_{i+d_{\rm e}+n,i+
d_{\rm e}+n+d_{\rm o}}^\dagger \mathcal{H}_{i,i+d_{\rm e}}^\pdag
\\
\label{eq:srep:f}
H_{2:2}^s(d_{\rm o},d_\text{e},d'_{\rm o})&:=  \sum_i \mathcal{S}_{i+d_\text{e},i+
d_\text{e}+d'_{o}}^\dagger \mathcal{S}_{i,i+d_{o}}^\pdag
\\
\label{eq:srep:g}
H_{2:2}^t(d_{\rm o},d_\text{e},d'_{\rm o})&:=  \sum_i \!\sum_{p=x,y,z}\!
\mathcal{T}_{i+d_\text{e},i+d_\text{e}+d'_{o}}^{\dagger \, p} 
\mathcal{T}_{i,i+d_{o}}^p
\end{align}
\eseq
The ``$\cdots$'' in Eq.\ \reqn{eq:j1j2_srep} denotes 3- and higher spinon interactions 
which may occur. But we neglect such contributions here, i.e., we develop 
an approach which is valid for low densities of spinons, but which may fail
in regimes where their density is higher.
 
The variable $n$ in the third line of Eq.\ \reqn{eq:j1j2_srep} takes odd values 
if $n>0$ and takes even values if $n<-d_{\rm o}$. 
We notice that $H_{3:1}^\vpdag(d_{e},n,d_{o})=0$ for $-d_\text{o}\leq n \leq0$. 
The term $H_{p:q}^\vpdag$ in Eqs.\ \reqn{eq:srep} is bilinear if $p+q=2$ and 
it is quartic if $p+q=4$. 
  
The so far unkown prefactors $\mathcal{A}$ in the ansatz \reqn{eq:j1j2_srep} are determined 
from the matrix elements of the Hamiltonian calculated in the orthonormal basis.
The following relations are useful to analyze the action of \reqn{eq:j1j2} 
on spinon  states
\bseq
\label{eq:iden}
\begin{align}
\label{eq:iden:a}
\left( 
\bm{S}_{(1,x)}^\vpdag \cdot \bm{S}_{(2,y)}^\vpdag \mp \frac{1}{4}
\right)
&\includegraphics[trim=7 18 0 0,scale=0.5]{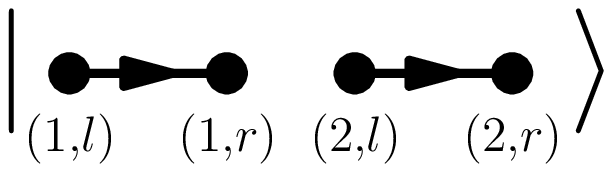}
\nn \\ 
& = \pm 
 \frac{1}{2} \includegraphics[trim=-0 18 0 0,scale=0.5]{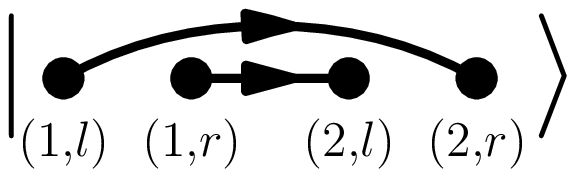}
\\
\label{eq:iden:b}
\left( 
\bm{S}_{(1,x)}^\vpdag \cdot \bm{S}_{(2)}^\vpdag \mp \frac{1}{4}
\right)
&\includegraphics[trim=7 23 0 0,scale=0.5]{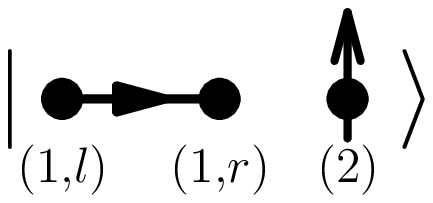}
\nn \\
&= \pm 
 \frac{1}{2} \includegraphics[trim=0 23 0 0,scale=0.5]{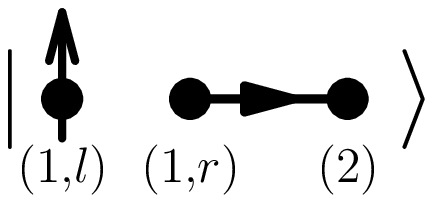}
\end{align}
\eseq
where $x,y \in \{\text{l},\text{r}\}$ and accordingly the 
spin operators $\bm{S}_{(1,x)}$ and $\bm{S}_{(2,y)}$ 
act on the lattice positions in the ket states in \reqn{eq:iden}. 
The upper sign in Eq.\ \reqn{eq:iden:a}  holds if 
$x$ and $y$ are different and the lower sign holds if they are the same. 
In Eq.\ \reqn{eq:iden:b}, the upper sign holds for $x=r$ and the lower sign for $x=l$. 
In addition to Eqs.\ \reqn{eq:expanI}, we use the following expansions for the 
3-spinon states
\bseq 
\label{eq:expanII}
\begin{align}
\label{eq:expanII:a}
\ket{\phi_{i}^\sigma \phi_{i+n,i+n+d}^{s}}
&=\frac{\sqrt{3}}{2} \left(\! -\frac{1}{2} \right)^{\!\!\frac{d-1}{2}}
\!\!\sum_{m_\text{e}\geq0} \left(\! -\frac{1}{2} \right)^{\!\!\frac{m_\text{e}}{2}} 
\ket{\Phi_{i-m_\text{e}}^{\sigma}}
\nn \\ & \hspace{-2.5cm}
+ \frac{\sqrt{3}}{2} \left(\! -\frac{1}{2} \right)^{\!\!\frac{n+1}{2}}
\!\sum_{m_\text{e}\geq0}^{d-3} \!\left(\! -\frac{1}{2} \right)^{\!\!\frac{m_\text{e}}{2}} 
\ket{\Phi_{i+n+d-m_\text{e}}^{\sigma}}
+\cdots,
\\
\label{eq:expanII:b}
\ket{ \phi_{i,i+d}^{s} \phi_{i+d+n}^\sigma }
&=\frac{\sqrt{3}}{2} \left(\! -\frac{1}{2} \right)^{\!\!\frac{n+1}{2}}
\!\!\sum_{m_\text{e}\geq0} \left(\! -\frac{1}{2} \right)^{\!\!\frac{m_\text{e}}{2}} 
\ket{\Phi_{i-m_\text{e}}^{\sigma}}
\nn \\ & \hspace{-2.5cm}
+ \frac{\sqrt{3}}{2} \left( \!-\frac{1}{2} \right)^{\!\!\frac{d-1}{2}}
\!\sum_{m_\text{e}\geq0}^{n-1} \!\left(\! -\frac{1}{2} \right)^{\!\!\frac{m_\text{e}}{2}} 
\ket{\Phi_{i+n+d-m_\text{e}}^{\sigma}}
+\cdots, 
\end{align}
\eseq
where ``$\cdots$'' stands for orthonormal 3-spinon states.

The prefactors $\mathcal{A}_{0:0}^\vpdag$ and $\mathcal{A}_{2:0}^\vpdag(d)$ can be computed 
by applying the Hamiltonian \reqn{eq:j1j2} to the vacuum \reqn{eq:spinon0}
and comparing the result with the application of the ansatz  \reqn{eq:j1j2_srep}
to the vacuum. In this way, we obtain
\begin{align}
\label{eq:H_on_spinon0}
\frac{H}{J_1}&\ket{0}= -\frac{3L}{8}\ket{0}
+\frac{\talpha}{2}\sum_{i\in {\rm even}} \left( \frac{1}{2}\ket{0}+ \ket{\phi_{i,i+3}^{s}} \right)
\nn \\ &
=-\frac{3L}{8} \ket{0} 
+ \frac{3\talpha}{2} \sum_{d_{\rm o}\geq 3} \sum_{i\in {\rm even}} 
\!\left(\!-\frac{1}{2} \right)^{\!\!\frac{d_{\rm o}+1}{2}} \!\ket{\Phi_{i,i+d_{\rm o}}^s}
\end{align}
where $\talpha:=1-2\alpha$. The second equality is derived using Eq.\ \reqn{eq:expanI:b} for $d=3$. 
The sum over $i$  runs over the even sites supposing that the first lattice site 
in the spinon vacuum \reqn{eq:spinon0} is even. For the other degenerate vacuum 
one sums over the odd sites. We should {\it not}, however, restrict the sum over $i$ 
in Eq.\ \reqn{eq:srep:d} to even (or odd) values only. 
Note, that both kinds of vacuum can be present on distinct pieces of the chain,
 separated by a spinon or any odd number of spinons as domain-walls. 
Thus it is important to keep the processes creating nested singlets. 
From Eq.\ \reqn{eq:H_on_spinon0}, we identify 
\bseq
\label{eq:A00_A20}
\begin{align}
\label{eq:A00}
\mathcal{A}_{0:0}^\vpdag&=-\frac{3}{8}J_1^\vpdag, \\
\label{eq:A20}
\mathcal{A}_{2:0}^\vpdag(d)&=\frac{3\talpha}{2}\left(-\frac{1}{2} 
\right)^{\!\!\frac{d+1}{2}} J_1^\vpdag.
\end{align}
\eseq
This result shows that the spin Hamiltonian \reqn{eq:j1j2} is not fully local represented in
spinon-pair operators. But the Bogoliubov 
prefactors \reqn{eq:A20} decrease exponentially upon increasing the distance $d$. 
Moreover, we see that there is no term linking the 0-spinon state to 4-spinon states. 
Such terms appear upon inclusion of longer-range (beyond NNN sites) 
spin-spin interactions in  \reqn{eq:j1j2}. 
At the MG point $\alpha=\frac{1}{2}$, all the Bogoliubov terms vanish indicating that the 
spinon vacuum \reqn{eq:spinon0} is an exact state of the system as one 
knows previously \cite{Majumdar1969b}.

To compute the hopping prefactors $\mathcal{A}_{1:1}^\vpdag(d)$ we apply the 
Hamiltonian \reqn{eq:j1j2} to the 1-spinon state $\ket{\Phi_{i}^{\sigma}}$
and compare it with the action of the ansatz \reqn{eq:j1j2_srep} on the same state.
Some lengthy calculations yield
\begin{align}
\frac{H}{J_1} \ket{\Phi_{i}^{\sigma}} = &
+\left( -\frac{3}{8}L+ \frac{5+2\talpha}{8}  \right) \ket{\Phi_{i}^{\sigma}}
\nn \\ &
+\left( \frac{2+\talpha}{8}  \right) \left( \ket{\Phi_{i+2}^{\sigma}}+ 
\ket{\Phi_{i-2}^{\sigma}} \right) 
\nn \\ &
-\frac{3\talpha}{4} \sum_{d_{e}\geq 4} \left( -\frac{1}{2} \right)^{\!\!\frac{d_{e}}{2}}
\left( \ket{\Phi_{i+d_{e}}^{\sigma}}+ \ket{\Phi_{i-d_{e}}^{\sigma}} \right) 
\nn \\ &
+ \cdots,
\end{align}
where ``$\cdots$'' denotes the 3-spinon contributions orthonormal to the 
1-spinon subspace which we neglect.
To derive this relation we employed Eqs.\ \reqn{eq:expanII} 
for $d=3$. No term linking the 1-spinon to the 5-spinon sector
is produced. After subtracting the contribution $\mathcal{A}_{0:0}^\vpdag$, 
the hopping prefactors read 
\bseq
\label{eq:A11}
\begin{align}
\mathcal{A}_{1:1}^\pdag(0) &= \frac{5+2\talpha}{8} J_1^\vpdag, \\
\mathcal{A}_{1:1}^\pdag(2) &= \frac{2+\talpha}{8} J_1^\vpdag, \\
\mathcal{A}_{1:1}^\pdag(d_\text{e}^\vpdag) &= 
-\frac{3\talpha}{4}\left( -\frac{1}{2} \right)^{\!\!\frac{d_\text{e}^\vpdag}{2}}  J_1^\vpdag \quad ; \quad d_\text{e}^\vpdag\geq4.
\end{align}
\eseq
At the MG point, $\talpha=0$, only hopping over two sites is present. However, there are 
long-range hopping processes present away from the MG point. The hopping prefactors 
decay exponentially with distance. 

It is more cumbersome to calculate the interaction coefficients 
$\mathcal{A}_{3:1}^\vpdag$, $\mathcal{A}_{2:2}^s$, and $\mathcal{A}_{2:2}^t$, 
analytically. 
The linked cluster expansion theorem allows us to determine 
the irreducible matrix elements $\mathcal{A}$ in Eq.\ \reqn{eq:j1j2_srep} 
on finite, but large enough clusters, in the thermodynamic limit. 
We refer the reader to Ref.\ \onlinecite{Knetter2003b} for details. 
We notice that the interaction coefficients can be computed {\it exactly} 
on finite clusters due to the locality of the spin Hamiltonian \reqn{eq:j1j2}.
 
We implemented a program to compute the Hamiltonian 
matrix elements between orthonormal spinon states. 
In the choice of the size of the cluster, one must keep in mind 
that each orthonormal spinon is extended over two sites to the left. 
Careful attention is also required in identifying the interaction coefficients 
$\mathcal{A}_{3:1}^\vpdag$ because of the finite overlap \reqn{eq:overlap_spinon3} 
between orthonormal 3-spinon states.
The interaction coefficients are calculated and reported in App.\ \ref{ap:sec:int_corff}. 
Even at the MG point the spin Hamiltonian \reqn{eq:j1j2} is 
not local in the spinon-pair operator representation, but the  non-local
terms fall off quickly with distance.

\section{Continuous unitary transformations}
\label{sec:cut}

We use continuous unitary transformations (CUT) to 
analyze the Hamiltonian \reqn{eq:j1j2_srep}. We briefly introduce the CUT 
method of which the first versions were suggested over 20 years ago 
\cite{Wegner1994}. 
The CUT method maps a given initial Hamiltonian $H$
to a final effective one by a unitary transformation which is
parametrized by an auxiliary parameter $\ell$ \cite{Kehrein2006}. 
The transformation is such that at $\ell=0$ the transformed Hamiltonian $H(\ell)$ 
equals the initial Hamiltonian $H$ and at $\ell=\infty$ the desired effective 
Hamiltonian is reached. The Hamiltonian during the flow is given by 
\be 
\label{eq:FE}
\partial_\ell^\vpdag H(\ell)=\left[ \eta(\ell), H(\ell) \right]
\ee
which is called the flow equation. The anti-hermitian operator $\eta(\ell)$, 
called generator, determines the essence of the transformation. 
One can use the Wegner generator \cite{Wegner1994}, 
the particle-conserving (pc) generator \cite{Stein1997,Mielke1998,Knetter2000}, or various 
reduced generators \cite{Fischer2010} to fully or partially rotate away 
the off-diagonal elements. 

By writing the initial Hamiltonian in a quasiparticle (QP) representation and eliminating the terms which change the number of QPs in the system, 
one can map a many-particle problem to a few-particle problem using the CUT method. 
There are different approximations to truncate the flow equation \reqn{eq:FE} so
that this systems of differential equations is closed. One may achieve
perturbative \cite{Knetter2000,Krull2012} or 
renormalized effective Hamiltonians \cite{Krull2012,Yang2011}. 
The method is applied to describe systems 
with elementary excitations such as triplons with spin $S=1$ 
\cite{Knetter2000a}, magnons \cite{Powalski2015}, 
electrons and holes \cite{Heidbrink2002b,Hafez2010b} and so on.

In the following, we employ the pc generator and keep in the transformed Hamiltonian $H(\ell)$ 
terms which are at most quartic. This truncation can be justified by two arguments.
The first argument refers to the density of spinons because the hexatic terms need at least
three spinons to become active. Thus at low concentrations of spinons
the bilinear terms acting on single spinons and the quartic terms
acting on pairs of spinons are the most important ones. 
The second argument refers to the scaling of the terms. For fermionic
and bosonic systems in one dimension it is known, that the bilinear and 
the quartic terms have the lowest scaling dimension \cite{kehre99,Powalski2015}. 
If the bilinear terms
are without mass term they are of equal scaling dimension, hence
of equal importance. Hexatic and higher terms are irrelevant in the
scaling sense.

In the bilinear approximation, the terms \reqn{eq:srep:a} to \reqn{eq:srep:d} define the operator basis. 
In the quartic approximation, we consider the terms \reqn{eq:srep:a} to 
\reqn{eq:srep:f}. For technical simplicity, we neglect to include the 
triplet 2-spinon interaction \reqn{eq:srep:g} because its treatment in the present
formalism breaks the translational symmetry of the lattice. 
Such a translational symmetry breaking is expected since we have implicitly chosen 
one of the two degenerate vacua.
One notices that no term linking the 0-spinon sector to the 4-spinon sector 
appears during the flow because the pc generator preserves the band-diagonal 
structure of the initial Hamiltonian \cite{Mielke1998, Knetter2000}.

To determine the set of flow equations we need to calculate the commutators between 
the operators of the basis \reqn{eq:srep} \cite{Krull2012}.
The necessary relations for the commutators up to quartic level are provided in 
Sect.\ \ref{sec:pair-operators} and in  Apps.\ \ref{ap:comm} and \ref{ap:HH}. 
While the flow equations up to bilinear level can be easily obtained 
it is a very tedious task to obtain the quartic contributions. 
This motivates us to implement a program for subsequent applications. 

To integrate the derived flow equations we limit the range of processes in real space
such that the distance between initial spinons, the distance between final spinons, 
and  the maximum distance between initial and final spinons is not larger than a 
maximum distance, $d_{\rm max}$.
The flow equations are integrated numerically with the initial conditions 
\reqn{eq:A00_A20}, \reqn{eq:A11}, \reqn{eq:A31_final}, and \reqn{eq:A22s_final}.

\section{Results}
\label{sec:res}

Before presenting the CUT results let us consider the simplest approximation, i.e.,
neglecting all off-diagonal terms, i.e., all terms which change the number
of spinons in \reqn{eq:j1j2_srep}. 
This approximation is expected to be most accurate at and near the MG point where 
the Bogoliubov  terms \reqn{eq:A20} vanish. But we stress that it is an
approximation even at the MG point because  terms exist which link
the 1-spinon states to the 3-spinons states.

The approximate 1-spinon dispersion neglecting all 
spinon-number changing terms reads
\begin{align}
\label{eq:disp_var}
\omega(k)&=\mac{A}_{1:1}^\vpdag(0)+2\sum_{d_\text{e}=2}\mac{A}_{1:1}
\vpdag(d_\text{e}) \cos(d_\text{e}^\vpdag k)
\nn \\
&= \frac{\alpha}{4} \left( 5+4\cos(2k) \right)J_1 
+ \widetilde{\alpha} \left( \frac{3}{8} + \frac{4+5\cos(2k)}{5+4\cos(2k)} 
\right)J_1,
\end{align}
based on Eqs.\ \reqn{eq:A11}.
This simple approximation recovers the variational results obtained by 
Brehmer et al.\ \cite{Brehmer1998,Lavarelo2014}. 
\emph{The formalism presented in this article allows us to systematically improve these 
variational results by rotating away the off-diagonal elements using the CUT or by other 
approaches.}

\begin{figure}[htb]
  \centering
  \includegraphics[width=0.78\columnwidth,angle=-90]{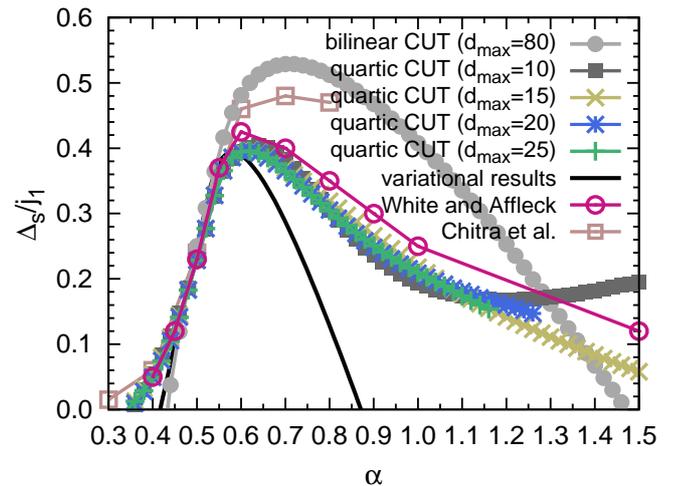}
  \caption{(Color online) The spin gap $\Delta_s$ versus the frustration $\alpha$.
	The various CUT truncation schemes are compared to the variational results by 
	Brehmer et al.\ \cite{Brehmer1998} and to the DMRG results by Chitra et al.\ 
	\cite{Chitra1995} and White and Affleck \cite{White1996}.}
  \label{fig:spin_gap}
\end{figure}

We define the spin gap $\Delta_s$ as twice the minimum energy of the spinon dispersion. 
Thus, the spin gap equals with the minimum of the 2-spinon continuum. 
We notice that the spinons are asymptotically free 
although they cannot pass each other. 
The spin gap $\Delta_s$ versus the frustration degree $\alpha$ 
is depicted in Fig.\ \ref{fig:spin_gap}.
The variational result \reqn{eq:disp_var} and the CUT results in the bilinear and quartic 
approximation for different maximum distances, $d_{\rm max}$, are shown. The 
density-matrix renormalization group (DMRG) results by Chitra et al.\ \cite{Chitra1995} 
and by White and Affleck \cite{White1996} are included for comparision. 

Fig.\ \ref{fig:spin_gap} shows that the spurious transition point predicted by the 
variational result at $\alpha \approx 0.87$ is shifted in CUT on bilinear level
 to $\alpha \approx 1.47$ making the spontaneously dimerized phase more stable
as expected from the DMRG results.
Both the variational result and the result from bilinear CUT vanish almost linearly at 
$\alpha \approx 0.43$. They do not show evidence of the exponentially slow vanishing
gap at the Berezinskii-Kosterlitz-Thouless (BKT) transition at $\alpha\approx0.241$.

The CUT results on the quartic level in Fig.\ \ref{fig:spin_gap} nicely capture the qualitatively behavior of the spin gap predicted by White and Affleck \cite{White1996}. 
The agreement between the two approaches is quantitative near the MG point. 
Lack of convergence of the flow equations 
prevents us to reach beyond $\alpha=1.26$ for 
$d_{\rm max}=20$ and beyond $\alpha=1.14$ for  $d_{\rm max}=25$.
The CUT on quartic level predicts the transition to the quasi-long-range spin liquid phase at 
$\alpha \approx 0.34$. This value is larger than the established critical value 
$\alpha_c \approx 0.241$. We believe that this 
deviation is due to the very slow vanishing of the spin gap near the BKT transition point 
\cite{Sorensen1998} which is only detectable if extremely long-range
processes are kept track of reliably.
Based on large-scale DMRG results \cite{Sorensen1998} 
the value of the spin gap at $\alpha=0.34$ is only about $0.004J_1$ 
which is too small to be detected in our approach.

\begin{figure}[htb]
  \centering
  \includegraphics[width=0.8\columnwidth,angle=-90]{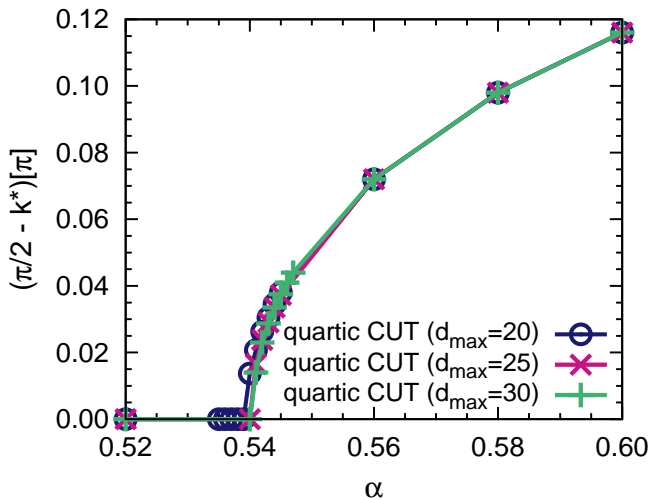}
  \caption{(Color online) The momentum $(\frac{\pi}{2}-k^*)$ versus the frustration
	$\alpha$; $k^*$ is the momentum at which the minimum of the spinon dispersion is located.
}
  \label{fig:k_gap}
\end{figure}

We denote by  $k^*$ the momentum at which the minimum of the spinon dispersion occurs. 
For weak frustration it occurs at $\pi/2$. Fig.\ \ref{fig:k_gap} denotes the deviation
$(\frac{\pi}{2}-k^*)$ versus the frustration $\alpha$.  
The minimum of the spinon dispersion moves from $k=\frac{\pi}{2}$ to an incommensurate 
value beyond the Lifshitz transition point $\alpha_l^\vpdag$. 
We find $\alpha_l^\vpdag=0.539$ for $d_{\rm max}=20$ and 
$\alpha_l^\vpdag=0.540$ for $d_{\rm max}=25$ and $d_{\rm max}=30$. 
This finding agrees quantitatively with the DMRG prediction 
$\alpha_l^\vpdag=0.538(1)$ by Deschner and S{\o}rensen \cite{Deschner2013}. 
However, it is difficult to capture the behavior of $k^*$ for $\alpha>0.538(1)$ 
by DMRG due to a plethora of level crossings \cite{Deschner2013}.
We expect $(\frac{\pi}{2}-k^*)\rightarrow \frac{\pi}{4}$ in the limit 
$\alpha \rightarrow \infty$ because the system approaches to weakly coupled
penetrating Heisenberg chains with lattice constant $2a$ so that the minimum
occurs for $a=1$ at $\frac{\pi}{4}$.

\begin{figure}[htb]
  \centering
  \includegraphics[width=1.15\columnwidth,angle=-90]{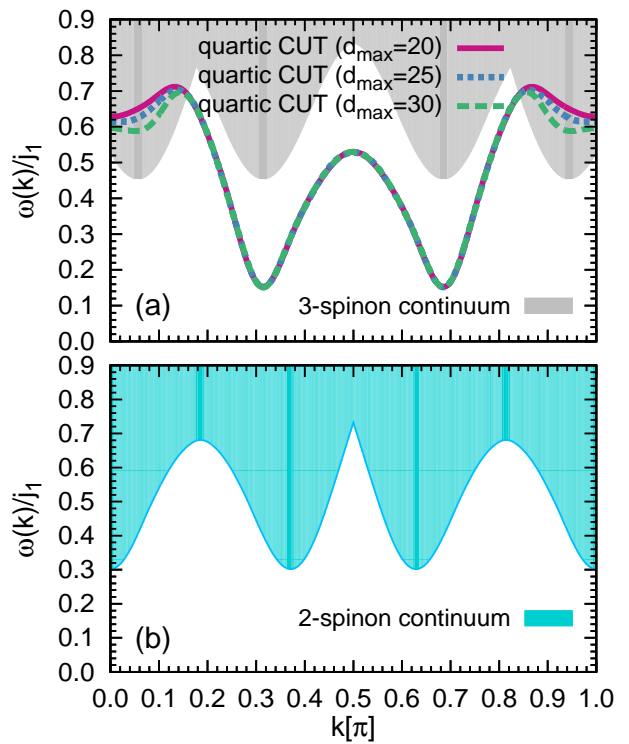}
  \caption{(Color online) The low-energy spectrum of the $J_1$-$J_2$ Heisenberg model 
	\reqn{eq:j1j2} with (a) an odd and (b) with an even number of sites at
 frustration $\alpha=0.8$. Panel (a) displays the 1-spinon dispersion and the 3-spinon continuum and panel (b) denotes the 2-spinon continuum. 
The continua are constructed by energy and momentum conservation
from the spinon dispersion with $d_{\rm max}=30$. The results 
for $d_{\rm max}=25$ are the same within line width.}
  \label{fig:disp}
\end{figure}

In Fig.\ \ref{fig:disp} we plot the low-energy spectrum of the $J_1$-$J_2$ 
Heisenberg model \reqn{eq:j1j2} with (a) an odd  and (b) an even number of sites. 
The value of the frustration $\alpha$ is set to $0.8$. 
Fig.\ \ref{fig:disp}a shows the 1-spinon dispersion and the 
3-spinon continuum while Fig.\ \ref{fig:disp}b shows the 2-spinon continuum. We used 
the spinon dispersion for $d_{\rm max}=30$ to construct the 2- and the 3-spinon continua
relying on energy and momentum conservation. The numerical results for $d_{\rm max}=25$
are essentially the same.
The 1-spinon dispersions obtained from different values of $d_{\rm max}$  deviate 
noticeably only inside the 3-spinon continuum where a clear distinction between 
1-spinon and 3-spinon states is not possible.
One can extend the CUT approach to describe the decay of the 1-spinon dispersion 
into the 3-spinon continuum in Fig.\ \ref{fig:disp}a \cite{Fischer2010}. 
But such an analysis is beyond the scope of the present article.  

We recall that the triplet 2-spinon interactions \reqn{eq:srep:g} are 
 neglected in the present treatment. These terms 
are present even in the initial Hamiltonian, see Eq.\ \reqn{eq:A22t_final}, and 
could make the CUT results on the quartic level even more accurate.

\section{Conclusions and Outlook}
\label{sec:conclusio}

The aim of the present paper was to provide the proof-of-principle
that models in terms of massive spinons can be formulated
for generic Hamiltonian, not only at special points such 
as the Majumdar-Ghosh point. This aim has been successfully 
realized. The formulation can be achieved in second quantization 
in the sense that the resulting effective Hamiltonian also 
applies to finite densities of spinons. 
The asset of such a formulation is that
the subsequent treatment can employ all methods known for such 
problems.

We have constructed an orthonormal spinon basis and introduced 
string spinon-pair operators which can capture the fractional nature 
of the spinon excitation and describe different spinon processes. 
By applying the spinon-pair operators on the spinon vacuum (1-spinon states) 
one can fully construct the $S=0$ and the $S=1$ Hilbert spaces of 
an even (odd) size lattice. This enables us to write spin Hamiltonians 
in spinon-pair representation in second quantization. 
Our representation is valid for low densities of spinons because it treats 
processes involving single spinons and pairs of them exactly into account. 
Only on the level of three or more spinons processes are neglected.

Here we used continuous unitary transformations to analyze the
second quantized Hamiltonians obtained in the first step. We showed that 
processes which change the number
of spinons can be systematically treated by this approach. 
They can be eliminated (``rotated away'') while their renormalization
of the properties of the elementary spinons is kept. In this way, we
obtained results for the spinon dispersion including gap and
incommensurability for the frustrated spin chain in a wide range
of frustration, i.e., not only for the Majumdar-Ghosh point where
the next-nearest neighbor interaction takes half the value of the
nearest-neighbor interaction.
The results are significantly improved over a purely variational
treatment. This illustrates the potential of the pursued approach.

Further research to establish effective models
in terms of their elementary excitations is called for.
One promising route of research consists in passing from 
treatments in real space to treatments in momentum space. 
Another current challenge
is to transfer the presented ideas from one to higher dimensions.

\begin{acknowledgments}
We are grateful to the Deutsche Forschungsgemeinschaft and the 
 Russian Foundation of Basic Research for support through the  ICRC TRR 160.
\end{acknowledgments}

\clearpage
\newpage

\onecolumngrid
\appendix

\section{Orthonormal 3-spinon states}
\label{ap:spinon3}

The orthonormal $S_{\rm tot}=\frac{1}{2}$ 3-spinon states 
$\ket{\Phi_{i}^{\sigma} \Phi_{i+d_1,i+d_1+d_2}^{s}}$ 
and $\ket{\Phi_{i,i+d_2}^{s}\Phi_{i+d_1+d_2}^{\sigma}}$ for $d_1 \geq 3$ 
are given by 
\bseq
\label{eq:spinon3_orth}
\begin{align}
\ket{\Phi_{i}^{\sigma} \Phi_{i+d_1,i+d_1+d_2}^{s}}&=\ket{\Phi_{i}^{\sigma}} \otimes 
\ket{\Phi_{i+d_1,i+d_1+d_2}^{s}},  \\
\ket{\Phi_{i,i+d_2}^{s}\Phi_{i+d_1+d_2}^{\sigma}} &=\ket{\Phi_{i,i+d_2}^{s}} \otimes 
\ket{\Phi_{i+d_1+d_2}^{\sigma}},
\end{align}
\eseq
where the orthonormal 1-spinon state $\ket{\Phi_{i}^{\sigma}}$ is defined by Eq.\ 
\reqn{eq:spinon1_l} and the orthonormal singlet 2-spinon state $\ket{\Phi_{i,i+d}^{s}}$ 
by Eq.\ \reqn{eq:spinon2_s_orth}.
If $d_1=1$, the orthonormal 3-spinon states 
\bseq
\label{eq:spinon3_distorted_scheme}
\begin{align}
\ket{\Phi_{i}^{\sigma} \Phi_{i+1,i+1+d}^{s}}
&=\includegraphics[trim=-0 24 0 0,scale=0.48]{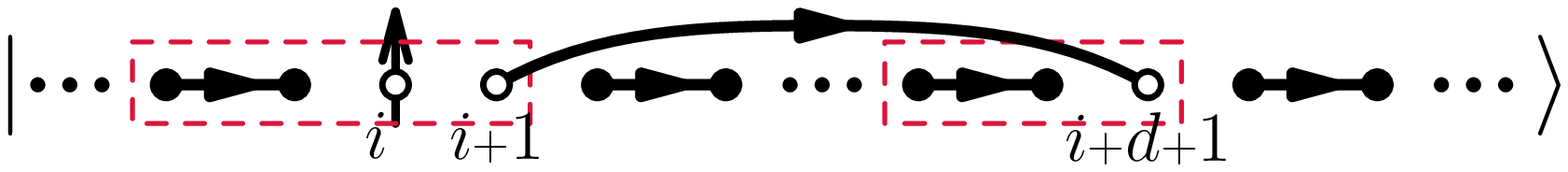}
\end{align}
and
\begin{align}
\ket{\Phi_{i,i+d}^{s}\Phi_{i+d+1}^{\sigma}}
&=\includegraphics[trim=-0 22 0 0,scale=0.48]{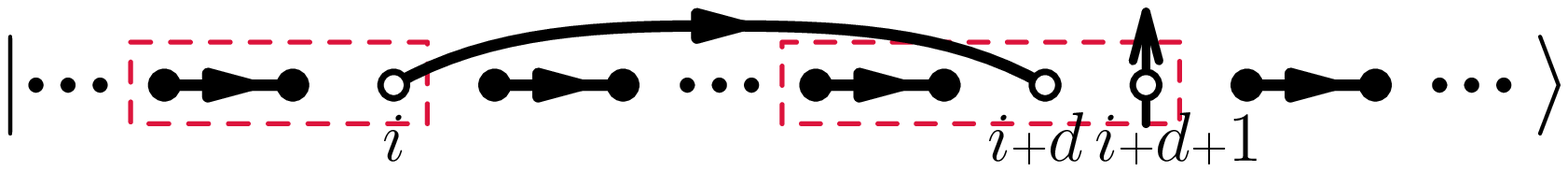} 
\end{align}
\eseq
are distorted. The empty circles indicate orthonormal spinons. 
We require orthonormality to derive these states. 
It is a subtle issue to identify the initial form 
of the distorted 3-spinon states. But it can be found in a systematic way. 
Each orthonormal spinon in Eqs.\ \reqn{eq:spinon3_distorted_scheme} 
has an extension of two sites to the left. 
This implies that 7 sites of the lattice are involved 
overall in 3-spinon states, see the dashed boxes in Eqs.\ 
\reqn{eq:spinon3_distorted_scheme}. On these 7 sites there are 12 states with 
$S_{\rm tot}=\frac{1}{2}$.  If $d>3$ in Eqs.\ \reqn{eq:spinon3_distorted_scheme}, 
2 of these 12 states are 5-spinon states which do not contribute to the orthonormal 3-spinon
states. We expect the same relation to be valid also 
for $d=3$. Hence, we arrive at the following ansatz with 10 parameters for the 
distorted 3-spinon states
\bseq
\label{eq:spinon3_dist}
\begin{alignat}{2}
\label{eq:spinon3_sus0}
\ket{\Phi_{i}^{\sigma} \Phi_{i+1,i+d+1}^{s}}
=&\frac{1}{N_1} 
\Big{(} 
\ket{\phi_{i}^{\sigma} \phi_{i+1,i+d+1}^{s}} +
\alpha_1 \ket{\phi_{i}^{\sigma} \phi_{i+1,i+d-1}^{s}} +
\alpha_2 \ket{\phi_{i-2}^{\sigma} \phi_{i+1,i+d+1}^{s}} +
\alpha_3 \ket{\phi_{i-2}^{\sigma} \phi_{i+1,i+d-1}^{s}}
\nn \\ 
&+ \alpha_4 \ket{\phi_{i-2}^{\sigma} \phi_{i-1,i+d+1}^{s}}
+\alpha_5 \ket{\phi_{i-2}^{\sigma} \phi_{i-1,i+d-1}^{s}}
+ \alpha_6 \ket{\phi_{i+d+1}^{\sigma}}
+\alpha_7 \ket{\phi_{i+d-1}^{\sigma}}
\nn \\
&+ \alpha_8 \ket{ \phi_{i-2,i+1}^{s} \phi_{i+d+1}^{\sigma} }
+\alpha_9 \ket{ \phi_{i-2,i+1}^{s} \phi_{i+d-1}^{\sigma} }
\Big{)},
\\ 
\label{eq:spinon3_s0su} 
\ket{ \Phi_{i,i+d}^{s} \Phi_{i+d+1}^{\sigma} }
=&\frac{1}{N_2} 
\Big{(} 
\ket{ \phi_{i,i+d}^{s} \phi_{i+d+1}^{\sigma} } +
\beta_1 \ket{ \phi_{i-2,i+d}^{s} \phi_{i+d+1}^{\sigma} } +
\beta_2 \ket{ \phi_{i,i+d-2}^{s} \phi_{i+d+1}^{\sigma} } +
\beta_3 \ket{ \phi_{i-2,i+d-2}^{s} \phi_{i+d+1}^{\sigma} }
\nn \\ 
&+\beta_4 \ket{ \phi_{i,i+d-2}^{s} \phi_{i+d-1}^{\sigma} } +
\beta_5 \ket{ \phi_{i-2,i+d-2}^{s} \phi_{i+d-1}^{\sigma} }
+ \beta_6 \ket{\phi_{i}^{\sigma} }
+\beta_7 \ket{\phi_{i-2}^{\sigma} }
\nn \\
&+\beta_8 \ket{\phi_{i}^{\sigma} \phi_{i+d-2,i+d}^{s} }
+\beta_9 \ket{\phi_{i-2}^{\sigma} \phi_{i+d-2,i+d}^{s} }
\Big{)}.
\end{alignat}
\eseq

The unknown prefactors are determined in an orthonormalization process
similar to the Gram-Schmidt algorithm. We start with 
the state $d=3$ in Eq.\ \reqn{eq:spinon3_sus0} and make it orthogonal 
to the 1-spinon states \reqn{eq:spinon1_l} and to the 3-spinon states 
\reqn{eq:spinon3_orth}.
It is too tedious a task to calculate the overlaps by hand. Hence we implemented 
a C$++$ program for this purpose. In this way, we obtained the eight independent 
equations
\allowdisplaybreaks
\bseq
\label{eq:overlap_sus0}
\begin{align}
\label{eq:overlap_sus0:a}
 \braket{\Phi_{i-2}^{\sigma}}{\Phi_{i}^{\sigma} \Phi_{i+1,i+4}^{s}}
&=0 \Longrightarrow -4\,\alpha_1^\pdag-4\,\alpha_2^\pdag+8\,\alpha_3^\pdag
+2\,\alpha_4^\pdag-4\,\alpha_5^\pdag
-1\,\alpha_6^\pdag+2\,\alpha_7^\pdag+2\,\alpha_8^\pdag-4\,\alpha_9^\pdag+2=0 
\\
 \braket{\Phi_{i}^{\sigma}}{\Phi_{i}^{\sigma} \Phi_{i+1,i+4}^{s}}
&=0 \Longrightarrow +4\,\alpha_1^\pdag+1\,\alpha_6^\pdag-2\,\alpha_7^\pdag-2=0 
\\
 \braket{\Phi_{i+2}^{\sigma}}{\Phi_{i}^{\sigma} \Phi_{i+1,i+4}^{s}}
&=0 \Longrightarrow +1\,\alpha^\pdag_4-2\,\alpha^\pdag_5-2\,\alpha^\pdag_6+4\,
\alpha^\pdag_7
+1\,\alpha^\pdag_8-2\,\alpha_9^\pdag=0 \\
 \braket{\Phi_{i+4}^{\sigma}}{\Phi_{i}^{\sigma} \Phi_{i+1,i+4}^{s}}
&=0 \Longrightarrow +1\,\alpha^\pdag_2-2\,\alpha^\pdag_4+4\,\alpha^\pdag_6-2\,
\alpha^\pdag_8-2=0 
\\
\label{eq:overlap_sus0:e}
 \braket{\Phi_{i-4}^{\sigma} \Phi_{i-1,i+2}^{s} }{\Phi_{i}^{\sigma} \Phi_{i+1,i+4}^{s}}
&=0 \Longrightarrow +3\,\alpha^\pdag_4-6\,\alpha^\pdag_5-1\,
\alpha^\pdag_8+2\,\alpha^\pdag_9=0 
\\
\label{eq:overlap_sus0:f}
 \braket{\Phi_{i-2}^{\sigma} \Phi_{i+1,i+4}^{s} }{\Phi_{i}^{\sigma} \Phi_{i+1,i+4}^{s}}
&=0 \Longrightarrow +2\,\alpha^\pdag_2-1\,\alpha^\pdag_8-1=0 
\\
\label{eq:overlap_sus0:g}
 \braket{\Phi_{i-4}^{\sigma} \Phi_{i-1,i+4}^{s} }{\Phi_{i}^{\sigma} \Phi_{i+1,i+4}^{s}}
&=0 \Longrightarrow +1\,\alpha^\pdag_2-2\,\alpha^\pdag_4=0 
\\
\label{eq:overlap_sus0:h}
 \braket{ \Phi_{i-2,i+1}^{s} \Phi_{i+4}^{\sigma} }{\Phi_{i}^{\sigma} \Phi_{i+1,i+4}^{s}}
&=0 \Longrightarrow +1\,\alpha^\pdag_2-2\,\alpha^\pdag_8=0 .
\end{align}
\eseq

Fulfilling the single equation \reqn{eq:overlap_sus0:a} is sufficient to make 
the state $\ket{\Phi_{i}^{\sigma} \Phi_{i+1,i+4}^{s}}$ orthogonal 
to all 1-spinon states $\ket{\Phi_{j}^{\sigma}}$ with $j \leq i-2$. 
The 1-spinon states $\ket{\Phi_{j}^{\sigma}}$ with $j \geq i+6$  
are trivially orthogonal to $\ket{\Phi_{i}^{\sigma} \Phi_{i+1,i+4}^{s}}$ 
leading to no condition for the prefactors $\alpha^\pdag_1 \cdots \alpha^\pdag_9$.
The overlaps $ \braket{\Phi_{i-1-d}^{\sigma} \Phi_{i-1,i+2}^{s} }{\Phi_{i}^{\sigma} 
\Phi_{i+1,i+4}^{s}}$ for $d=3,5,\cdots$ all vanish based on 
Eq.\ \reqn{eq:overlap_sus0:e}. Similarly each of the 
Eqs.\ \reqn{eq:overlap_sus0:f}, \reqn{eq:overlap_sus0:g}, and 
\reqn{eq:overlap_sus0:h} orthogonalize the distorted state $\ket{\Phi_{i}^{\sigma} 
\Phi_{i+1,i+4}^{s}}$ to a class of orthonormal 3-spinon states.

From Eqs.\ \reqn{eq:overlap_sus0} we find
\be 
\label{eq:spinon3_sus0_pre}
\alpha_1^\pdag =\frac{4}{9}+\frac{\alpha_9^\pdag}{3}; \, 
\alpha_2^\pdag =\frac{2}{3}; \, 
\alpha_3^\pdag =\frac{2}{9}+\frac{2\, \alpha_9^\pdag}{3}; \, 
\alpha_4^\pdag =\frac{1}{3}; \, 
\alpha_5^\pdag =\frac{1}{9}+\frac{\alpha_9^\pdag}{3}; \, 
\alpha_6^\pdag =\frac{2}{3}; \, 
\alpha_7^\pdag =\frac{2}{9}+\frac{2\, \alpha_9^\pdag}{3}; \, 
\alpha_8^\pdag =\frac{1}{3}; \, 
\ee
where $\alpha_9^\pdag$ is left undetermined. In addition to the relations 
\reqn{eq:spinon3_sus0_pre}, the orthogonalization process of 
the distorted state \reqn{eq:spinon3_sus0} with $d=5$ requires
to satisfy
\be 
 \braket{ \Phi_{i-2,i+1}^{s} \Phi_{i+4}^{\sigma} }{\Phi_{i}^{\sigma} \Phi_{i+1,i+6}^{s}}
= +1\,\alpha^\pdag_2-2\,\alpha^\pdag_3-2\,\alpha^\pdag_8+4\,\alpha^\pdag_9=0
\ee
leading to 
\be 
\alpha_9^\pdag = \frac{1}{6}; \, N_1^\pdag =\frac{\sqrt{3}}{2\, \sqrt{2}};
\ee
where $N_1^\pdag$ is the normalizing prefactor. We checked that the states 
\reqn{eq:spinon3_sus0} for different $i$ and $d$ are also orthogonal. 

The prefactors $\beta_1^\pdag \cdots \beta_9^\pdag$ in Eq.\ \reqn{eq:spinon3_s0su} 
can be computed similarly.
We orthogonalize the distorted state \reqn{eq:spinon3_s0su} to the 1-spinon state 
\reqn{eq:spinon1_l} and to the 3-spinon states \reqn{eq:spinon3_orth} and 
\reqn{eq:spinon3_sus0}. For $d=3$, this yields the eight independent 
equations
\bseq
\label{eq:overlap_s0su}
\begin{align}
\label{eq:overlap_s0su:a}
 \braket{\Phi_{i-2}^{\sigma}}{ \Phi_{i,i+3}^{s} \Phi_{i+4}^{\sigma} }=0 
& \Longrightarrow  -4\,\beta_1^\pdag-1\,\beta_2^\pdag+2\,\beta_3^\pdag+2\,
\beta_4^\pdag-4\,\beta_5^\pdag
-4\,\beta_6^\pdag+8\,\beta_7^\pdag+2\,\beta_8^\pdag-4\,\beta_9^\pdag+2=0 
\\
 \braket{\Phi_{i}^{\sigma}}{ \Phi_{i,i+3}^{s} \Phi_{i+4}^{\sigma} } =0
& \Longrightarrow +1\,\beta_2^\pdag-2\,\beta_4^\pdag+4\,\beta_6^\pdag-2\,\beta_8^\pdag-2=0 
\\
 \braket{\Phi_{i+2}^{\sigma}}{ \Phi_{i,i+3}^{s} \Phi_{i+4}^{\sigma} } =0
& \Longrightarrow -2\,\beta^\pdag_2+1\,\beta^\pdag_3+4\,\beta^\pdag_4-2\,\beta^\pdag_5=0 
\\
 \braket{\Phi_{i+4}^{\sigma}}{ \Phi_{i,i+3}^{s} \Phi_{i+4}^{\sigma} } =0
& \Longrightarrow +1\,\beta^\pdag_1+4\,\beta^\pdag_2-2\,
\beta^\pdag_3-2\,\beta^\pdag_8+1\,\beta^\pdag_9-2=0 
\\
\label{eq:overlap_s0su:e}
 \braket{ \Phi_{i-2,i+1}^{s} \Phi_{i+4}^{\sigma} }{ \Phi_{i,i+3}^{s} \Phi_{i+4}^{\sigma} } =0
& \Longrightarrow -1\,\beta^\pdag_1+2\,\beta^\pdag_3-1\,\beta^\pdag_9=0 \\
\label{eq:overlap_s0su:f}
 \braket{\Phi_{i-4}^{\sigma} \Phi_{i-1,i+2}^{s} }{ \Phi_{i,i+3}^{s} \Phi_{i+4}^{\sigma} } =0
& \Longrightarrow +1\,\beta^\pdag_3-2\,\beta^\pdag_5=0 \\
\label{eq:overlap_s0su:g}
 \braket{\Phi_{i-2}^{\sigma} \Phi_{i+1,i+4}^{s} }{ \Phi_{i,i+3}^{s} \Phi_{i+4}^{\sigma} } =0
& \Longrightarrow +3\,\beta^\pdag_3+3\,\beta^\pdag_8-6\,\beta^\pdag_9-1=0 \\
\label{eq:overlap_s0su:h}
 \braket{ \Phi_{i-4}^{\sigma} \Phi_{i-1,i+4}^{s} }{ \Phi_{i,i+3}^{s} \Phi_{i+4}^{\sigma} } =0
& \Longrightarrow +1\,\beta^\pdag_1-3\,\beta^\pdag_9=0
\end{align}
\eseq
leading to
\be 
\label{eq:spinon3_s0su_pre}
\beta^\pdag_1 = 3\, \beta^\pdag_9;\,  \beta^\pdag_2 = \frac{2}{3};\,  
\beta^\pdag_3 = 2\,\beta^\pdag_9;\,  
\beta^\pdag_4 = \frac{1}{3};\,  \beta^\pdag_5 = \beta^\pdag_9;\,  
\beta^\pdag_6 = \frac{2}{3};\,  
\beta^\pdag_7 = 2\,\beta^\pdag_9;\,  \beta^\pdag_8 = \frac{1}{3} \quad.
\ee
The prefactor $\beta_9^\pdag$ remains unspecified. 


The orthogonalization process for the state 
\reqn{eq:spinon3_s0su} 
with $d=5$ leads to the same results as \reqn{eq:spinon3_s0su_pre}. 
However, one still needs to orthogonalize 
the distorted states \reqn{eq:spinon3_s0su} for different $i$ and $d$. 
We find the equation
\be 
 \braket{ \Phi_{i,i+5}^{s} \Phi_{i+6}^{\sigma} }{ \Phi_{i+2,i+5}^{s} \Phi_{i+6}^{\sigma} } =0
\Longrightarrow +18\,\beta^2_9-15\,\beta^\pdag_9+2=0 \quad,
\ee
where we used the relations \reqn{eq:spinon3_s0su_pre}. This gives us the two solutions 
$\beta_9^\pdag=\frac{1}{6}$ and $\frac{2}{3}$. We choose the solution 
$\beta_9^\pdag=\frac{1}{6}$ as it indicates that the spinon 
at position $i$ is of type {\it left}. The other solution $\beta_9^\pdag=\frac{2}{3}$ 
corresponds to a spinon of type {\it right} at $i-2$. Hence, we obtain
\bearr
\alpha^\pdag_1&=&\beta^\pdag_1 = \frac{1}{2};\,  \alpha^\pdag_2=\beta^\pdag_2 = \frac{2}{3};\, 
\alpha^\pdag_3=\beta^\pdag_3 = \frac{1}{3};\, \alpha^\pdag_4=\beta^\pdag_4 = \frac{1}{3};\, 
\alpha^\pdag_5=\beta^\pdag_5 = \frac{1}{6};\,  \alpha^\pdag_6=\beta^\pdag_6 = \frac{2}{3};\, \nn \\
\alpha^\pdag_7&=&\beta^\pdag_7 = \frac{1}{3};\,  \alpha^\pdag_8=\beta^\pdag_8 = \frac{1}{3};\,
\alpha^\pdag_9=\beta^\pdag_9 = \frac{1}{6};\,  N_1^\pdag=N_2^\pdag = 
\frac{\sqrt{3}}{2\,\sqrt{2}} \quad.
\eearr
This concludes the construction of the distorted orthonormal 
3-spinon  states \reqn{eq:spinon3_sus0} and \reqn{eq:spinon3_s0su}.

\section{Commutators}
\label{ap:comm}

The commutators between two neutral hopping operators are given by
\bseq
\label{eq:comm_HH}
\begin{alignat}{2}
\label{eq:comm_HpHp}
[ \mathcal{H}_{i,i+d}^{\vphantom{\dagger}} , \mathcal{H}_{j,j+d'}^{\vphantom{\dagger}} ] =&
(\delta_{i,j+d'}^{\vphantom{\dagger}} \mathcal{H}_{i-d',i+d}^{\vphantom{\dagger}} 
-\delta_{j,i+d}^{\vphantom{\dagger}} \mathcal{H}_{i,i+d+d'}^{\vphantom{\dagger}} )
+(  \delta_{i+d,j-1}^{\vphantom{\dagger}} \!+ \delta_{i-1,j+d'}^{\vphantom{\dagger}} ) 
( \mathcal{H}_{i,i+d}^{\vphantom{\dagger}} \mathcal{H}_{j,j+d'}^{\vphantom{\dagger}} 
-\mathcal{H}_{j,j+d'}^{\vphantom{\dagger}} \mathcal{H}_{i,i+d}^{\vphantom{\dagger}} ) 
 \nn \\&
+ \left( \theta(j-i) \theta(i+d-j)  \theta(j+d'-i-d) \mathcal{H}_{i,i+d}^\vpdag 
\mathcal{H}_{j,j+d'}^\vpdag -
\left\{
\begin{aligned}
i & \rightarrow j \\
d & \rightarrow d'
\end{aligned}
\right\}
\right)
\\
\label{eq:comm_HnHp}
[ \mathcal{H}_{i+d,i}^\vpdag , \mathcal{H}_{j,j+d'}^\vpdag ] =& 
\delta_{i+d,j+d'}^\vpdag 
\left(  
\theta(d-d'+1) \mathcal{P}_{i+d-d'+1,i+d}^\vpdag +
\theta(d'-d) \mathcal{P}_{i+1,i+d}^\vpdag
\right) 
\mathcal{H}_{i+d-d',i}^\vpdag
\nn \\ &
-\delta_{i,j}^\vpdag 
\left(  
\theta(d'-d+1) \mathcal{P}_{i,i+d-1}^\vpdag 
+\theta(d-d') \mathcal{P}_{i,i+d'-1}^\vpdag
\right) 
\mathcal{H}_{i+d,i+d'}^\vpdag 
\end{alignat}
\eseq
with $d$ and $d'$ positive. We defined $\mathcal{P}_{i,i-1}^{\vphantom{\dagger}}=1$.
If the hopping operator brings two spinons in the singlet state to nearest-neighbor sites 
the result is zero. This explains why the second contribution in Eq.\ \reqn{eq:comm_HpHp} 
can lead to a finite value. These commutators 
are required only up to bilinear level due to the approximation that we employ 
in this paper. Hence, Eqs. \reqn{eq:comm_HH} simplify to
\be 
[ \mathcal{H}_{i,i+d}^\vpdag , \mathcal{H}_{j,j+d'}^\vpdag ] = 
(\delta_{i,j+d'}^{\vphantom{\dagger}} \mathcal{H}_{i-d',i+d}^{\vphantom{\dagger}} 
-\delta_{j,i+d}^{\vphantom{\dagger}} \mathcal{H}_{i,i+d+d'}^{\vphantom{\dagger}} )
+\cdots
\ee
with $d' \geq 0$ and arbitrary $d$. 
Hermitian conjugation yields the commutator for $d' \leq 0$.

Two creation singlet operators do not necessarily commute; we find
\be
\label{eq:comm_SdSd}
\left[ \mathcal{S}_{i,i+d}^{\dagger}, \mathcal{S}_{j,j+d'}^{\dagger}  \right] 
= \mathcal{S}_{i,i+d}^{\dagger} \mathcal{S}_{j,j+d'}^{\dagger} \theta(i-j)  \theta(j+d'-i-d)
- \mathcal{S}_{j,j+d'}^{\dagger} \mathcal{S}_{i,i+d}^{\dagger} \theta(j-i)  \theta(i+d-j-d').
\ee
This reflects the fact that in the  creation of nested singlets 
the order of the singlet operators matters. 

\section{Expansion of products of hopping operators}
\label{ap:HH}

The product of two neutral hopping operators can be expanded as
\begin{align}
 \mac{H}_{i+n,i+n+d'}^\vpdag &\mac{H}_{i,i+d}^\vpdag= \delta_{n,d}^\vpdag
\left( 
\mac{H}_{i,i+d+d'}^\vpdag 
\vphantom{
- \sum_{m_\text{o}\geq1} \theta(d-m_\text{o}^\vpdag) \theta(m_\text{o}^\vpdag-d-d')
\left(
 \mathcal{S}_{i+d+d',i+m_\text{o}^\vpdag+2}^{\dagger} 
\mathcal{S}_{i,i+m_\text{o}^\vpdag+2}^\vpdag
+\!\!\!\sum_{p=x,y,z}\!\!  \mathcal{T}_{i+d+d',i+m_\text{o}^\vpdag}^{p\,\dagger} 
\mathcal{T}_{i,i+m_\text{o}^\vpdag}^p
\right)
}
\right. \nn \\ & \left.
- \sum_{m_\text{o}\geq1} \theta(-m_\text{o}^\vpdag+d) \theta(m_\text{o}^\vpdag-d-d')
\left(
 \mathcal{S}_{i+d+d',i+m_\text{o}^\vpdag+2}^{\dagger} 
\mathcal{S}_{i,i+m_\text{o}^\vpdag+2}^\vpdag
+\!\!\!\sum_{p=x,y,z}\!\!  \mathcal{T}_{i+d+d',i+m_\text{o}^\vpdag}^{p\,\dagger} 
\mathcal{T}_{i,i+m_\text{o}^\vpdag}^p
\right)
\right. \nn \\ & \left.
- \sum_{m_\text{o}\geq1} \theta(-m_\text{o}^\vpdag-d) \theta(m_\text{o}^\vpdag+d+d')
\left(
 \mathcal{S}_{i-m_\text{o}^\vpdag-2,i+d+d'}^{\dagger} 
\mathcal{S}_{i-m_\text{o}^\vpdag-2,i}^\vpdag
+\!\!\!\sum_{p=x,y,z}\!\!  \mathcal{T}_{i-m_\text{o}^\vpdag,i+d+d'}^{p\,\dagger} 
\mathcal{T}_{i-m_\text{o}^\vpdag,i}^p
\right)
\right)
\nn \\ & \hspace{-1.5cm}
+\theta(+n\!-\!2)\theta(n\!-\!d\!-\!2)\theta(n\!+\!d'\!-\!d\!-\!2)
 \mathcal{S}_{i+d,i+n+d'}^{\dagger} \mathcal{S}_{i,i+n}^\vpdag
+\theta(+n)\theta(n\!-\!d)\theta(n\!+\!d'\!-\!d)
\!\!\!\sum_{p=x,y,z}\!\!  \mathcal{T}_{i+d,i+n+d'}^{p\,\dagger} \mathcal{T}_{i,i+n}^p
\nn \\ & \hspace{-1.5cm}
+\theta(-n\!-\!2)\theta(d\!-\!n\!-\!2)\theta(d\!-\!n\!-\!d'\!-\!2)
 \mathcal{S}_{i+n+d',i+d}^{\dagger} \mathcal{S}_{i+n,i}^\vpdag
+\theta(-n)\theta(d\!-\!n)\theta(d\!-\!n\!-\!d')
\!\!\!\sum_{p=x,y,z}\!\!  \mathcal{T}_{i+n+d',i+d}^{p\,\dagger} \mathcal{T}_{i+n,i}^p 
\nn \\ & \hspace{-1.5cm} 
+\cdots
\end{align}
where 3- and higher spinon interactions are neglected.

\section{Interaction coefficients}
\label{ap:sec:int_corff}

In order to compute the interaction coefficients $\mathcal{A}_{3:1}$ one has to take 
into account the possible overlap \reqn{eq:overlap_spinon3} between orthonormal 3-spinon states. 
For each $S_{\rm tot}=\frac{1}{2}$ orthonormal 3-spinon state 
$\ket{\Phi_{i}^\sigma\Phi_{i+n,i+n+d}^s}$ we define a dual state 
given by
\be 
\ket{\Phi_{i}^\sigma\Phi_{i+n,i+n+d}^s}^{(d)}:=
\begin{cases}
\ket{\Phi_{i,i+n}^s\Phi_{i+n+d}^\sigma}& \quad {\rm if}\quad n>0
\\
\ket{\Phi_{i+n}^\sigma\Phi_{i+n+d,i}^s}& \quad {\rm if}\quad n<-d
\end{cases}
\ee
Each orthonormal 3-spinon state is orthogonal to all other states except to its dual state. 
The overlap is given by
\be 
\braket{\Phi_{i}^\sigma\Phi_{i+n,i+n+d}^s}{\Phi_{i}^\sigma\Phi_{i+n,i+n+d}^s}^{\!(d)}=
-\frac{1}{2}.
\ee
We notice that there is no such finite overlap for the distorted 3-spinon states 
\reqn{eq:spinon3_dist}. Hence they do not have a dual state.

We define the Hamiltonian matrix element 
\be 
\mac{C}_{3:1}^\vpdag(d_\text{e},n,d_\text{o}):=\bra{\Phi_{i+d_\text{e}}^\sigma
\Phi_{i+d_\text{e}+n,i+d_\text{e}+n+d_\text{o}}^s}H 
\ket{\Phi_{i}^\sigma}
\ee 
and its corresponding dual
\be 
\mac{C}_{3:1}^{(d)}(d_\text{e},n,d_\text{o}):=\vphantom{\Phi}^{(d)}\!\!
\bra{\Phi_{i+d_\text{e}}^\sigma
\Phi_{i+d_\text{e}+n,i+d_\text{e}+n+d_\text{o}}^s}H \ket{\Phi_{i}^\sigma}.
\ee 
One has the relation 
\be 
\mac{C}_{3:1}^{(d)}(d_\text{e},n,d_\text{o})=
\begin{cases}
 \mac{C}_{3:1}^{\vpdag}(d_\text{e}+n+d_\text{o},-n-d_\text{o},n) \quad &{\rm if} \quad n>0 
\\
 \mac{C}_{3:1}^{\vpdag}(d_\text{e}+n,d_\text{o},-n-d_\text{o}) \quad &{\rm if} 
\quad n+d_\text{o}<0.
\end{cases}
\ee
Similarly, we define the dual of the interaction coefficient 
$\mac{A}_{3:1}^{\vpdag}(d_\text{e},n,d_\text{o})$ as
\be 
\mac{A}_{3:1}^{(d)}(d_\text{e},n,d_\text{o}):=
\begin{cases}
 \mac{A}_{3:1}^{\vpdag}(d_\text{e}+n+d_\text{o},-n-d_\text{o},n) \quad &{\rm if} \quad n>0 
\\
 \mac{A}_{3:1}^{\vpdag}(d_\text{e}+n,d_\text{o},-n-d_\text{o}) \quad &{\rm if} 
\quad n+d_\text{o}<0
\end{cases}
\ee
which links to the 1-spinon state 
$\ket{\Phi_{i}^\sigma}$ to $\ket{\Phi_{i+d_\text{e}}^\sigma
\Phi_{i+d_\text{e}+n,i+d_\text{e}+n+d_\text{o}}^s}^{(d)}$.

Using the relation \reqn{eq:j1j2_srep}, the Hamiltonian matrix elements 
$\mac{C}_{3:1}^{\vpdag}(d_\text{e},n,d_\text{o})$ and 
$\mac{C}_{3:1}^{(d)}(d_\text{e},n,d_\text{o})$ are given by 
\bseq 
\label{eq:C_31}
\begin{align}
\mac{C}_{3:1}^{\vpdag}(d_\text{e},n,d_\text{o})&\!=\!\mac{A}_{3:1}^{\vpdag}(d_\text{e},n,d_\text{o}) 
-\frac{1}{2}\mac{A}_{3:1}^{(d)}(d_\text{e},n,d_\text{o})
+\delta_{n,0}^\vpdag \mac{A}_{2:0}^{\vpdag}(d_\text{o})
-\frac{1}{2} \delta_{n+d_\text{o},-d_\text{e}}^\vpdag \mac{A}_{2:0}^{\vpdag}(n)
-\frac{1}{2} \delta_{n,-d_\text{e}}^\vpdag \mac{A}_{2:0}^{\vpdag}(-n-d_\text{o})
\\
\mac{C}_{3:1}^{(d)}(d_\text{e},n,d_\text{o})&\!=\!\mac{A}_{3:1}^{(d)}(d_\text{e},n,d_\text{o}) 
-\frac{1}{2}\mac{A}_{3:1}^{\vpdag}(d_\text{e},n,d_\text{o})
-\frac{1}{2}\delta_{n,0}^\vpdag \mac{A}_{2:0}^{\vpdag}(d_\text{o})
+ \delta_{n+d_\text{o},-d_\text{e}}^\vpdag \mac{A}_{2:0}^{\vpdag}(n)
+ \delta_{n,-d_\text{e}}^\vpdag \mac{A}_{2:0}^{\vpdag}(-n-d_\text{o})
\end{align}
\eseq
where we supposed $\mac{A}_{2:0}^{\vpdag}(d)=0$ if $d$ is even or if $d<3$. In addition, we assumed $n>1$ or $n+d_\text{o}<-1$ which means that the final state is not distorted. 
From Eqs.\ \reqn{eq:C_31} we find 
\bseq
\label{eq:A_31}
\begin{align} 
\mac{A}_{3:1}^{\vpdag}(d_\text{e},n,d_\text{o}) =\frac{4}{3}\mac{C}_{3:1}^{\vpdag}(d_\text{e},n,d_\text{o})
+\frac{2}{3}\mac{C}_{3:1}^{(d)}(d_\text{e},n,d_\text{o})-\delta_{d_\text{e},0}^\vpdag 
\mac{A}_{2:0}^{\vpdag}(d_\text{o})
\quad ; \quad \{n\!>\!1 ~{\rm or}~ n\!<\!-1\!-\!d_\text{o}\}.
\end{align}

In the case that the final state is distorted, the interaction coefficient is given by 
\begin{align} 
\mac{A}_{3:1}^{\vpdag}(d_\text{e},n,d_\text{o}) =\mac{C}_{3:1}^{\vpdag}(d_\text{e},n,d_\text{o})
-\delta_{d_\text{e},0}^\vpdag \mac{A}_{2:0}^{\vpdag}(d_\text{o})
\quad ; \quad \{n=1 ~{\rm or}~ n+d_\text{o}=-1\}.
\end{align}
\eseq
Eqs.\ \reqn{eq:A_31} help to determine the interaction coefficients 
$\mac{A}_{3:1}^{\vpdag}(d_\text{e},n,d_\text{o})$ from the Hamiltonian 
matrix elements and the Bogoliubov prefactors. 
Since $\mac{H}_{3:1}^{\vpdag}(d_\text{e},n,d_\text{o})$ is cluster additive, we can 
compute $\mac{A}_{3:1}^{\vpdag}(d_\text{e},n,d_\text{o})$ for each specific 
$d_\text{e}$, $n$, and $d_\text{o}$ on finite clusters of sufficient size
\cite{Knetter2003b}. Comparing coefficients, we arrive at the analytical, 
compact form 
\bseq 
\label{eq:A31_final}
\begin{align}
\mathcal{A}_{3:1}^\pdag &(d_{e}^\vpdag,1,d_\text{o}^\vpdag) =  
\frac{\sqrt{2}}{12} \delta_{d_{o},3}^\pdag \left( (3\alpha-1) 
\delta_{d_{e}^\vpdag,-4}^\pdag+
2 \alpha \delta_{d_{e}^\vpdag,-2}^\pdag \right)
+ \left(-\frac{1}{2} \right)^{\frac{(d_\text{o}^\vpdag+3)}{2}} \left(  
(3-2\sqrt{2}) \delta_{d_{e}^\vpdag,0}^\pdag -\sqrt{2}\delta_{d_{e}^\vpdag,-2}^\pdag  \right) 
(1-2\alpha),
\\
\mathcal{A}_{3:1}^\pdag &(d_{e}^\vpdag,n,3) = 
\left(-\frac{1}{2} \right)^{\frac{(n+3)}{2}} 
\left( 
\frac{6\alpha-2}{3}  \delta_{-d_{e}^\vpdag,n+3}^\pdag
+\frac{3-2\alpha}{3}  \delta_{-d_{e}^\vpdag,n+1}^\pdag
+\frac{1-2\alpha}{2}  \delta_{-d_{e}^\vpdag,n-1}^\pdag
\right)
; \quad n=3,5,\cdots \quad,
\\
\mathcal{A}_{3:1}^\pdag &(d_{e}^\vpdag,n,d_\text{o}^\vpdag) = 
\left(-\frac{1}{2} \right)^{\frac{(n+d_\text{o}^\vpdag)}{2}} 
\left( 
\delta_{-d_{e}^\vpdag,n+1}^\pdag
+\frac{1}{2}  \delta_{-d_{e}^\vpdag,n-1}^\pdag
\right)
(1-2\alpha)
; \quad n=3,5,\cdots, \quad d_\text{o}^\vpdag=5,7,\cdots \quad,
\\
\mathcal{A}_{3:1}^\pdag &(d_{e}^\vpdag,-d_\text{o}^\vpdag\!-\!1,d_\text{o}^\vpdag) = 
\left(-\frac{1}{2} \right)^{\frac{(d_\text{o}^\vpdag+3)}{2}} 
\left( 
\frac{6-3\sqrt{2}}{2} (1-2\alpha) \delta_{d_{e}^\vpdag,0}^\pdag
-\frac{\sqrt{2}}{2} \delta_{d_{e}^\vpdag,0}^\pdag
-2\sqrt{2} \alpha \delta_{d_{e}^\vpdag,2}^\pdag
\right),
\\
\mathcal{A}_{3:1}^\pdag &(d_{e}^\vpdag,-d_\text{o}^\vpdag\!-\!3,d_\text{o}^\vpdag) = 
\left(-\frac{1}{2} \right)^{\frac{(d_\text{o}^\vpdag+3)}{2}} 
\left( 
\frac{3\alpha-1}{3} \delta_{d_{e}^\vpdag,0}^\pdag
+\frac{3-2\alpha}{6} \delta_{d_{e}^\vpdag,2}^\pdag
+(1-2\alpha) \delta_{d_{e}^\vpdag,4}^\pdag
\right),
\\
\mathcal{A}_{3:1}^\pdag &(d_{e}^\vpdag,n,d_\text{o}^\vpdag) = 
\left(-\frac{1}{2} \right)^{\frac{(d_{e}^\vpdag+d_\text{o}^\vpdag+1)}{2}} 
\left( 
\frac{1}{2} \delta_{d_{e}^\vpdag+1,-n-d_\text{o}^\vpdag}^\pdag
-2 \delta_{d_{e}^\vpdag-1,-n-d_\text{o}^\vpdag}^\pdag
\right) (1-2\alpha) \, ; 
\quad n=\!-d_\text{o}\vpdag \!-5,-d_\text{o}\vpdag \!-7,\cdots \, .
\end{align}
\eseq

The interaction coefficients $\mathcal{A}_{2:2}^s(d_{o}^\vpdag,d_\text{e}^\vpdag,d'_{o})$ 
and $\mathcal{A}_{2:2}^t(d_{o}^\vpdag,d_\text{e}^\vpdag,d'_{o})$ for specific $d_\text{o}$, 
$d_\text{e}$, and $d'_\text{o}$ 
can also be calculated on finite clusters \cite{Knetter2003b}. From the analysis of 
the coefficients we deduce
\bseq 
\label{eq:A22s_final}
\begin{align}
\mathcal{A}_{2:2}^s(d_\text{o},d_\text{e},3) &=  \frac{5-7\talpha}{24} \delta_{d_{o},3}^\vpdag\delta_{d_\text{e},0}^\vpdag+
\left( -\frac{1}{2} \right)^{\frac{d_{o}+3}{2}} 
\left( \delta_{d_{o}^\vpdag-3,d_\text{e}}(1-3\talpha)-\delta_{d_{o}^\vpdag-1,d_\text{e}}(1+\frac{\talpha}{2}) \right), 
\\
\mathcal{A}_{2:2}^s(d_{o}^\vpdag,d_\text{e},d'_{o}) 
&=  -\frac{3\talpha}{2} \left( -\frac{1}{2} \right)^{\frac{d_{o}+d'_{o}}{2}} \delta_{d_{o}^\vpdag-1,d_\text{e}} 
\quad ; \quad d'_{o}=5,7,\cdots \quad.
\end{align}
\eseq
where we supposed $d_\text{e}\geq 0$. 

The triplet channel interactions are given by
\bseq 
\label{eq:A22t_final}
\begin{align}
\mathcal{A}_{2:2}^t(d_\text{o},0,d'_\text{o}) =&  \left( 
\frac{23\alpha-10+4\sqrt{6}(1-2\alpha)}{6} \right) 
\delta_{d_{o},1}^\vpdag \delta_{d'_\text{o},1}^\vpdag
+\frac{\alpha}{2}\left( \frac{4\sqrt{6}}{9} - 1 \right) 
\left( \delta_{d_{o},1}^\vpdag \delta_{d'_\text{o},3}^\vpdag + \delta_{d_{o},3}^\vpdag 
\delta_{d'_\text{o},1}^\vpdag \right)
+\frac{(\alpha-1)}{18}\delta_{d_{o},3}^\vpdag \delta_{d'_\text{o},3}^\vpdag
\nn \\ 
&+\frac{(1-2\alpha)}{2}\left( \frac{4\sqrt{6}}{3} - 3 \right) \left( -\frac{1}{2}  
\right)^{\frac{d_\text{o}+d'_\text{o}}{2}}
\left( \delta_{d_\text{o},1}^\vpdag + \delta_{d'_\text{o},1}^\vpdag \right)
\\
\mathcal{A}_{2:2}^t(d_{o}^\vpdag,d_\text{e} ,1) 
=& +\frac{(2-\sqrt{6})\alpha }{6} \delta_{d_\text{o},1}^\vpdag 
\delta_{d_\text{e},2}^\vpdag
+\frac{\alpha}{2} \left( \frac{4\sqrt{6}}{9} -1 \right) 
\delta_{d_\text{o},3}^\vpdag \delta_{d_\text{e},2}^\vpdag
-\frac{\sqrt{6}\alpha}{3}\left( -\frac{1}{2} \right)^{\frac{d_\text{e}}{2}} 
\delta_{d_\text{o}+1,d_\text{e}}^\vpdag
\nn \\
&+\left( \frac{3(1-2\alpha)}{4} + \frac{(3\alpha-2)\sqrt{6}}{6}\right)
\left( -\frac{1}{2} \right)^{\frac{d_\text{e}}{2}} 
\delta_{d_\text{o}-1,d_\text{e}}^\vpdag \quad ; \quad d_\text{e}^\vpdag\geq \!2
\\
\mathcal{A}_{2:2}^t(d_{o}^\vpdag,d_\text{e} ,3) =& 
+\left( -\frac{1}{2} \right)^{\frac{d_\text{e}}{2}+2} 
\left(  
(1-2\alpha) \delta_{d_\text{o}+1,d_\text{e}}^\vpdag + \frac{3-2\alpha}{6} 
\delta_{d_\text{o}-1,d_\text{e}}^\vpdag 
+\frac{3\alpha-1}{3} \delta_{d_\text{o}-3,d_\text{e}}^\vpdag
\right) 
\nn \\ 
&+\delta_{d_\text{o},1}^\vpdag \delta_{d_\text{e},2}^\vpdag 
\frac{3-\sqrt{6}}{24}(1-2\alpha)
\quad ;\quad d_\text{e}^\vpdag \geq 2
\\
\mathcal{A}_{2:2}^t(d_{o}^\vpdag,d_\text{e} ,d'_{o}) =&
\frac{(1-2\alpha)}{2} \left( -\frac{1}{2} \right)^{\frac{d_\text{o}+d'_\text{o}}{2}}
\left( 
\delta_{d_\text{o}-1,d_\text{e}}^\vpdag - \delta_{d_\text{o}+1,d_\text{e}}^\vpdag + 
\frac{3-\sqrt{6}}{3} 
\delta_{d_\text{o},1}^\vpdag \delta_{d_\text{e},2}^\vpdag 
\right)
\quad ;\quad d_\text{e}^\vpdag \geq 2 \quad , \quad d'_\text{o} \geq 5 ~.
\end{align}
\eseq
The prefactors $\mathcal{A}_{2:2}^s(d_{o}^\vpdag,d_\text{e}^\vpdag,d'_{o})$ 
and $\mathcal{A}_{2:2}^t(d_{o}^\vpdag,d_\text{e}^\vpdag,d'_{o})$ with 
$d_\text{e}<0$ can be calculated from the symmetry relation 
\be 
\mathcal{A}_{2:2}^\vpdag(d_{o}^\pdag,d_\text{e}^\vpdag,d'_{o})=
\mathcal{A}_{2:2}^\vpdag(d'_{o},-d_\text{e}^\vpdag,d_{o}^\vpdag).
\ee

\twocolumngrid

\section*{References}

\begin{thebibliography}{65}%
\makeatletter
\providecommand \@ifxundefined [1]{%
 \@ifx{#1\undefined}
}%
\providecommand \@ifnum [1]{%
 \ifnum #1\expandafter \@firstoftwo
 \else \expandafter \@secondoftwo
 \fi
}%
\providecommand \@ifx [1]{%
 \ifx #1\expandafter \@firstoftwo
 \else \expandafter \@secondoftwo
 \fi
}%
\providecommand \natexlab [1]{#1}%
\providecommand \enquote  [1]{``#1''}%
\providecommand \bibnamefont  [1]{#1}%
\providecommand \bibfnamefont [1]{#1}%
\providecommand \citenamefont [1]{#1}%
\providecommand \href@noop [0]{\@secondoftwo}%
\providecommand \href [0]{\begingroup \@sanitize@url \@href}%
\providecommand \@href[1]{\@@startlink{#1}\@@href}%
\providecommand \@@href[1]{\endgroup#1\@@endlink}%
\providecommand \@sanitize@url [0]{\catcode `\\12\catcode `\$12\catcode
  `\&12\catcode `\#12\catcode `\^12\catcode `\_12\catcode `\%12\relax}%
\providecommand \@@startlink[1]{}%
\providecommand \@@endlink[0]{}%
\providecommand \url  [0]{\begingroup\@sanitize@url \@url }%
\providecommand \@url [1]{\endgroup\@href {#1}{\urlprefix }}%
\providecommand \urlprefix  [0]{URL }%
\providecommand \Eprint [0]{\href }%
\providecommand \doibase [0]{http://dx.doi.org/}%
\providecommand \selectlanguage [0]{\@gobble}%
\providecommand \bibinfo  [0]{\@secondoftwo}%
\providecommand \bibfield  [0]{\@secondoftwo}%
\providecommand \translation [1]{[#1]}%
\providecommand \BibitemOpen [0]{}%
\providecommand \bibitemStop [0]{}%
\providecommand \bibitemNoStop [0]{.\EOS\space}%
\providecommand \EOS [0]{\spacefactor3000\relax}%
\providecommand \BibitemShut  [1]{\csname bibitem#1\endcsname}%
\let\auto@bib@innerbib\@empty
\bibitem [{\citenamefont {Holstein}\ and\ \citenamefont
  {Primakoff}(1940)}]{holst40}%
  \BibitemOpen
  \bibfield  {author} {\bibinfo {author} {\bibfnamefont {T.}~\bibnamefont
  {Holstein}}\ and\ \bibinfo {author} {\bibfnamefont {H.}~\bibnamefont
  {Primakoff}},\ }\href@noop {} {\bibfield  {journal} {\bibinfo  {journal}
  {Phys. Rev.}\ }\textbf {\bibinfo {volume} {58}},\ \bibinfo {pages} {1098}
  (\bibinfo {year} {1940})}\BibitemShut {NoStop}%
\bibitem [{\citenamefont {Auerbach}(1994)}]{auerb94}%
  \BibitemOpen
  \bibfield  {author} {\bibinfo {author} {\bibfnamefont {A.}~\bibnamefont
  {Auerbach}},\ }\href@noop {} {\emph {\bibinfo {title} {Interacting Electrons
  and Quantum Magnetism}}},\ Graduate Texts in Contemporary Physics\ (\bibinfo
  {publisher} {Springer},\ \bibinfo {address} {New York},\ \bibinfo {year}
  {1994})\BibitemShut {NoStop}%
\bibitem [{\citenamefont {Sachdev}\ and\ \citenamefont
  {Bhatt}(1990)}]{Sachdev1990}%
  \BibitemOpen
  \bibfield  {author} {\bibinfo {author} {\bibfnamefont {S.}~\bibnamefont
  {Sachdev}}\ and\ \bibinfo {author} {\bibfnamefont {R.~N.}\ \bibnamefont
  {Bhatt}},\ }\href {\doibase 10.1103/PhysRevB.41.9323} {\bibfield  {journal}
  {\bibinfo  {journal} {Phys. Rev. B}\ }\textbf {\bibinfo {volume} {41}},\
  \bibinfo {pages} {9323} (\bibinfo {year} {1990})}\BibitemShut {NoStop}%
\bibitem [{\citenamefont {Uhrig}\ and\ \citenamefont
  {Schulz}(1996)}]{uhrig96b}%
  \BibitemOpen
  \bibfield  {author} {\bibinfo {author} {\bibfnamefont {G.~S.}\ \bibnamefont
  {Uhrig}}\ and\ \bibinfo {author} {\bibfnamefont {H.~J.}\ \bibnamefont
  {Schulz}},\ }\href@noop {} {\bibfield  {journal} {\bibinfo  {journal} {Phys.
  Rev. B}\ }\textbf {\bibinfo {volume} {54}},\ \bibinfo {pages} {9624(R)}
  (\bibinfo {year} {1996})}\BibitemShut {NoStop}%
\bibitem [{\citenamefont {Knetter}\ and\ \citenamefont
  {Uhrig}(2000)}]{Knetter2000}%
  \BibitemOpen
  \bibfield  {author} {\bibinfo {author} {\bibfnamefont {C.}~\bibnamefont
  {Knetter}}\ and\ \bibinfo {author} {\bibfnamefont {G.}~\bibnamefont
  {Uhrig}},\ }\href {\doibase 10.1007/s100510050026} {\bibfield  {journal}
  {\bibinfo  {journal} {The European Physical Journal B - Condensed Matter and
  Complex Systems}\ }\textbf {\bibinfo {volume} {13}},\ \bibinfo {pages} {209}
  (\bibinfo {year} {2000})}\BibitemShut {NoStop}%
\bibitem [{\citenamefont {Trebst}\ \emph {et~al.}(2000)\citenamefont {Trebst},
  \citenamefont {Monien}, \citenamefont {Hamer}, \citenamefont {Weihong},\ and\
  \citenamefont {Singh}}]{trebs00}%
  \BibitemOpen
  \bibfield  {author} {\bibinfo {author} {\bibfnamefont {S.}~\bibnamefont
  {Trebst}}, \bibinfo {author} {\bibfnamefont {H.}~\bibnamefont {Monien}},
  \bibinfo {author} {\bibfnamefont {C.~J.}\ \bibnamefont {Hamer}}, \bibinfo
  {author} {\bibfnamefont {Z.}~\bibnamefont {Weihong}}, \ and\ \bibinfo
  {author} {\bibfnamefont {R.~R.~P.}\ \bibnamefont {Singh}},\ }\href@noop {}
  {\bibfield  {journal} {\bibinfo  {journal} {Phys. Rev. Lett.}\ }\textbf
  {\bibinfo {volume} {85}},\ \bibinfo {pages} {4373} (\bibinfo {year}
  {2000})}\BibitemShut {NoStop}%
\bibitem [{\citenamefont {Sachdev}(1999)}]{sachd99}%
  \BibitemOpen
  \bibfield  {author} {\bibinfo {author} {\bibfnamefont {S.}~\bibnamefont
  {Sachdev}},\ }\href@noop {} {\emph {\bibinfo {title} {Quantum Phase
  Transitions}}}\ (\bibinfo  {publisher} {Cambridge University Press},\
  \bibinfo {address} {Cambridge, UK},\ \bibinfo {year} {1999})\BibitemShut
  {NoStop}%
\bibitem [{\citenamefont {Weihong}\ \emph {et~al.}(1999)\citenamefont
  {Weihong}, \citenamefont {Hamer},\ and\ \citenamefont {Oitmaa}}]{weiho99b}%
  \BibitemOpen
  \bibfield  {author} {\bibinfo {author} {\bibfnamefont {Z.}~\bibnamefont
  {Weihong}}, \bibinfo {author} {\bibfnamefont {C.~J.}\ \bibnamefont {Hamer}},
  \ and\ \bibinfo {author} {\bibfnamefont {J.}~\bibnamefont {Oitmaa}},\
  }\href@noop {} {\bibfield  {journal} {\bibinfo  {journal} {Phys. Rev. B}\
  }\textbf {\bibinfo {volume} {60}},\ \bibinfo {pages} {6608} (\bibinfo {year}
  {1999})}\BibitemShut {NoStop}%
\bibitem [{\citenamefont {Uhrig}\ \emph {et~al.}(2004)\citenamefont {Uhrig},
  \citenamefont {Schmidt},\ and\ \citenamefont {Gr\"uninger}}]{Uhrig2004}%
  \BibitemOpen
  \bibfield  {author} {\bibinfo {author} {\bibfnamefont {G.~S.}\ \bibnamefont
  {Uhrig}}, \bibinfo {author} {\bibfnamefont {K.~P.}\ \bibnamefont {Schmidt}},
  \ and\ \bibinfo {author} {\bibfnamefont {M.}~\bibnamefont {Gr\"uninger}},\
  }\href {\doibase 10.1103/PhysRevLett.93.267003} {\bibfield  {journal}
  {\bibinfo  {journal} {Phys. Rev. Lett.}\ }\textbf {\bibinfo {volume} {93}},\
  \bibinfo {pages} {267003} (\bibinfo {year} {2004})}\BibitemShut {NoStop}%
\bibitem [{\citenamefont {Takatsu}\ \emph {et~al.}(1997)\citenamefont
  {Takatsu}, \citenamefont {Shiramura},\ and\ \citenamefont
  {Tanaka}}]{takat97}%
  \BibitemOpen
  \bibfield  {author} {\bibinfo {author} {\bibfnamefont {K.-I.}\ \bibnamefont
  {Takatsu}}, \bibinfo {author} {\bibfnamefont {W.}~\bibnamefont {Shiramura}},
  \ and\ \bibinfo {author} {\bibfnamefont {H.}~\bibnamefont {Tanaka}},\
  }\href@noop {} {\bibfield  {journal} {\bibinfo  {journal} {J. Phys. Soc.
  Jap.}\ }\textbf {\bibinfo {volume} {66}},\ \bibinfo {pages} {1611} (\bibinfo
  {year} {1997})}\BibitemShut {NoStop}%
\bibitem [{\citenamefont {Matsumoto}\ \emph {et~al.}(2002)\citenamefont
  {Matsumoto}, \citenamefont {Normand}, \citenamefont {Rice},\ and\
  \citenamefont {Sigrist}}]{matsu02}%
  \BibitemOpen
  \bibfield  {author} {\bibinfo {author} {\bibfnamefont {M.}~\bibnamefont
  {Matsumoto}}, \bibinfo {author} {\bibfnamefont {B.}~\bibnamefont {Normand}},
  \bibinfo {author} {\bibfnamefont {T.~M.}\ \bibnamefont {Rice}}, \ and\
  \bibinfo {author} {\bibfnamefont {M.}~\bibnamefont {Sigrist}},\ }\href@noop
  {} {\bibfield  {journal} {\bibinfo  {journal} {Phys. Rev. Lett.}\ }\textbf
  {\bibinfo {volume} {89}},\ \bibinfo {pages} {077203} (\bibinfo {year}
  {2002})}\BibitemShut {NoStop}%
\bibitem [{\citenamefont {Nohadani}\ \emph {et~al.}(2004)\citenamefont
  {Nohadani}, \citenamefont {Wessel}, \citenamefont {Normand},\ and\
  \citenamefont {Haas}}]{nohad04}%
  \BibitemOpen
  \bibfield  {author} {\bibinfo {author} {\bibfnamefont {O.}~\bibnamefont
  {Nohadani}}, \bibinfo {author} {\bibfnamefont {S.}~\bibnamefont {Wessel}},
  \bibinfo {author} {\bibfnamefont {B.}~\bibnamefont {Normand}}, \ and\
  \bibinfo {author} {\bibfnamefont {S.}~\bibnamefont {Haas}},\ }\href@noop {}
  {\bibfield  {journal} {\bibinfo  {journal} {Phys. Rev. B}\ }\textbf {\bibinfo
  {volume} {69}},\ \bibinfo {pages} {220402(R)} (\bibinfo {year}
  {2004})}\BibitemShut {NoStop}%
\bibitem [{\citenamefont {Sirker}\ \emph {et~al.}(2005)\citenamefont {Sirker},
  \citenamefont {Wei\"s{}e},\ and\ \citenamefont {Sushkov}}]{sirke05}%
  \BibitemOpen
  \bibfield  {author} {\bibinfo {author} {\bibfnamefont {J.}~\bibnamefont
  {Sirker}}, \bibinfo {author} {\bibfnamefont {A.}~\bibnamefont {Wei\"s{}e}}, \
  and\ \bibinfo {author} {\bibfnamefont {O.~P.}\ \bibnamefont {Sushkov}},\
  }\href@noop {} {\bibfield  {journal} {\bibinfo  {journal} {J. Phys. Soc.
  Jap.}\ }\textbf {\bibinfo {volume} {74}},\ \bibinfo {pages} {Suppl.\ 129}
  (\bibinfo {year} {2005})}\BibitemShut {NoStop}%
\bibitem [{\citenamefont {Jensen}(2011)}]{jense11}%
  \BibitemOpen
  \bibfield  {author} {\bibinfo {author} {\bibfnamefont {J.}~\bibnamefont
  {Jensen}},\ }\href@noop {} {\bibfield  {journal} {\bibinfo  {journal} {Phys.
  Rev. B}\ }\textbf {\bibinfo {volume} {83}},\ \bibinfo {pages} {064420}
  (\bibinfo {year} {2011})}\BibitemShut {NoStop}%
\bibitem [{\citenamefont {Faddeev}\ and\ \citenamefont
  {Takhtajan}(1981)}]{Faddeev1981}%
  \BibitemOpen
  \bibfield  {author} {\bibinfo {author} {\bibfnamefont {L.}~\bibnamefont
  {Faddeev}}\ and\ \bibinfo {author} {\bibfnamefont {L.}~\bibnamefont
  {Takhtajan}},\ }\href {\doibase
  http://dx.doi.org/10.1016/0375-9601(81)90335-2} {\bibfield  {journal}
  {\bibinfo  {journal} {Physics Letters A}\ }\textbf {\bibinfo {volume} {85}},\
  \bibinfo {pages} {375 } (\bibinfo {year} {1981})}\BibitemShut {NoStop}%
\bibitem [{\citenamefont {Karbach}\ \emph {et~al.}(1997)\citenamefont
  {Karbach}, \citenamefont {M\"uller}, \citenamefont {Bougourzi}, \citenamefont
  {Fledderjohann},\ and\ \citenamefont {M\"utter}}]{Karbach1997}%
  \BibitemOpen
  \bibfield  {author} {\bibinfo {author} {\bibfnamefont {M.}~\bibnamefont
  {Karbach}}, \bibinfo {author} {\bibfnamefont {G.}~\bibnamefont {M\"uller}},
  \bibinfo {author} {\bibfnamefont {A.~H.}\ \bibnamefont {Bougourzi}}, \bibinfo
  {author} {\bibfnamefont {A.}~\bibnamefont {Fledderjohann}}, \ and\ \bibinfo
  {author} {\bibfnamefont {K.-H.}\ \bibnamefont {M\"utter}},\ }\href {\doibase
  10.1103/PhysRevB.55.12510} {\bibfield  {journal} {\bibinfo  {journal} {Phys.
  Rev. B}\ }\textbf {\bibinfo {volume} {55}},\ \bibinfo {pages} {12510}
  (\bibinfo {year} {1997})}\BibitemShut {NoStop}%
\bibitem [{\citenamefont {Caux}\ and\ \citenamefont {Hagemans}(2006)}]{caux06}%
  \BibitemOpen
  \bibfield  {author} {\bibinfo {author} {\bibfnamefont {J.-S.}\ \bibnamefont
  {Caux}}\ and\ \bibinfo {author} {\bibfnamefont {R.}~\bibnamefont
  {Hagemans}},\ }\href@noop {} {\bibfield  {journal} {\bibinfo  {journal} {J.
  Stat. Mech.: Theor. Exp.}\ ,\ \bibinfo {pages} {P12013}} (\bibinfo {year}
  {2006})}\BibitemShut {NoStop}%
\bibitem [{\citenamefont {Mourigal}\ \emph {et~al.}(2013)\citenamefont
  {Mourigal}, \citenamefont {Enderle}, \citenamefont {Kl\"opperpieper},
  \citenamefont {Caux}, \citenamefont {Stunault},\ and\ \citenamefont
  {R\protect{\o}nnow}}]{mouri13}%
  \BibitemOpen
  \bibfield  {author} {\bibinfo {author} {\bibfnamefont {M.}~\bibnamefont
  {Mourigal}}, \bibinfo {author} {\bibfnamefont {M.}~\bibnamefont {Enderle}},
  \bibinfo {author} {\bibfnamefont {A.}~\bibnamefont {Kl\"opperpieper}},
  \bibinfo {author} {\bibfnamefont {J.-S.}\ \bibnamefont {Caux}}, \bibinfo
  {author} {\bibfnamefont {A.}~\bibnamefont {Stunault}}, \ and\ \bibinfo
  {author} {\bibfnamefont {H.~M.}\ \bibnamefont {R\protect{\o}nnow}},\
  }\href@noop {} {\bibfield  {journal} {\bibinfo  {journal} {Nat. Phys.}\
  }\textbf {\bibinfo {volume} {9}},\ \bibinfo {pages} {435} (\bibinfo {year}
  {2013})}\BibitemShut {NoStop}%
\bibitem [{\citenamefont {Haldane}(1988)}]{halda88a}%
  \BibitemOpen
  \bibfield  {author} {\bibinfo {author} {\bibfnamefont {F.~D.~M.}\
  \bibnamefont {Haldane}},\ }\href@noop {} {\bibfield  {journal} {\bibinfo
  {journal} {Phys. Rev. Lett.}\ }\textbf {\bibinfo {volume} {60}},\ \bibinfo
  {pages} {635} (\bibinfo {year} {1988})}\BibitemShut {NoStop}%
\bibitem [{\citenamefont {Shastry}(1988)}]{shast88}%
  \BibitemOpen
  \bibfield  {author} {\bibinfo {author} {\bibfnamefont {B.~S.}\ \bibnamefont
  {Shastry}},\ }\href@noop {} {\bibfield  {journal} {\bibinfo  {journal} {Phys.
  Rev. Lett.}\ }\textbf {\bibinfo {volume} {60}},\ \bibinfo {pages} {639}
  (\bibinfo {year} {1988})}\BibitemShut {NoStop}%
\bibitem [{\citenamefont {Enderle}\ \emph {et~al.}(2010)\citenamefont
  {Enderle}, \citenamefont {F\o{a}k}, , \citenamefont {Mikeska}, \citenamefont
  {Kremer}, \citenamefont {Prokofiev},\ and\ \citenamefont {Assmus}}]{ender10}%
  \BibitemOpen
  \bibfield  {author} {\bibinfo {author} {\bibfnamefont {M.}~\bibnamefont
  {Enderle}}, \bibinfo {author} {\bibfnamefont {B.}~\bibnamefont {F\o{a}k}}, ,
  \bibinfo {author} {\bibfnamefont {H.-J.}\ \bibnamefont {Mikeska}}, \bibinfo
  {author} {\bibfnamefont {R.}~\bibnamefont {Kremer}}, \bibinfo {author}
  {\bibfnamefont {A.}~\bibnamefont {Prokofiev}}, \ and\ \bibinfo {author}
  {\bibfnamefont {W.}~\bibnamefont {Assmus}},\ }\href@noop {} {\bibfield
  {journal} {\bibinfo  {journal} {Phys. Rev. Lett.}\ }\textbf {\bibinfo
  {volume} {104}},\ \bibinfo {pages} {237207} (\bibinfo {year}
  {2010})}\BibitemShut {NoStop}%
\bibitem [{\citenamefont {Coldea}\ \emph {et~al.}(2001)\citenamefont {Coldea},
  \citenamefont {Tennant}, \citenamefont {Tsvelik},\ and\ \citenamefont
  {Tylczynski}}]{colde01a}%
  \BibitemOpen
  \bibfield  {author} {\bibinfo {author} {\bibfnamefont {R.}~\bibnamefont
  {Coldea}}, \bibinfo {author} {\bibfnamefont {D.~A.}\ \bibnamefont {Tennant}},
  \bibinfo {author} {\bibfnamefont {A.~M.}\ \bibnamefont {Tsvelik}}, \ and\
  \bibinfo {author} {\bibfnamefont {Z.}~\bibnamefont {Tylczynski}},\
  }\href@noop {} {\bibfield  {journal} {\bibinfo  {journal} {Phys. Rev. Lett.}\
  }\textbf {\bibinfo {volume} {86}},\ \bibinfo {pages} {1335} (\bibinfo {year}
  {2001})}\BibitemShut {NoStop}%
\bibitem [{\citenamefont {Bocquet}\ \emph {et~al.}(2001)\citenamefont
  {Bocquet}, \citenamefont {Essler}, \citenamefont {Tsvelik},\ and\
  \citenamefont {Gogolin}}]{bocqu01}%
  \BibitemOpen
  \bibfield  {author} {\bibinfo {author} {\bibfnamefont {M.}~\bibnamefont
  {Bocquet}}, \bibinfo {author} {\bibfnamefont {F.~H.~L.}\ \bibnamefont
  {Essler}}, \bibinfo {author} {\bibfnamefont {A.~M.}\ \bibnamefont {Tsvelik}},
  \ and\ \bibinfo {author} {\bibfnamefont {A.~O.}\ \bibnamefont {Gogolin}},\
  }\href@noop {} {\bibfield  {journal} {\bibinfo  {journal} {Phys. Rev. B}\
  }\textbf {\bibinfo {volume} {64}},\ \bibinfo {pages} {094425} (\bibinfo
  {year} {2001})}\BibitemShut {NoStop}%
\bibitem [{\citenamefont {Kohno}\ \emph {et~al.}(2007)\citenamefont {Kohno},
  \citenamefont {Starykh},\ and\ \citenamefont {Balents}}]{kohno07}%
  \BibitemOpen
  \bibfield  {author} {\bibinfo {author} {\bibfnamefont {M.}~\bibnamefont
  {Kohno}}, \bibinfo {author} {\bibfnamefont {O.~A.}\ \bibnamefont {Starykh}},
  \ and\ \bibinfo {author} {\bibfnamefont {L.}~\bibnamefont {Balents}},\
  }\href@noop {} {\bibfield  {journal} {\bibinfo  {journal} {Nat. Phys.}\
  }\textbf {\bibinfo {volume} {3}},\ \bibinfo {pages} {790} (\bibinfo {year}
  {2007})}\BibitemShut {NoStop}%
\bibitem [{\citenamefont {Shastry}\ and\ \citenamefont
  {Sutherland}(1981)}]{Shastry1981}%
  \BibitemOpen
  \bibfield  {author} {\bibinfo {author} {\bibfnamefont {B.~S.}\ \bibnamefont
  {Shastry}}\ and\ \bibinfo {author} {\bibfnamefont {B.}~\bibnamefont
  {Sutherland}},\ }\href {\doibase 10.1103/PhysRevLett.47.964} {\bibfield
  {journal} {\bibinfo  {journal} {Phys. Rev. Lett.}\ }\textbf {\bibinfo
  {volume} {47}},\ \bibinfo {pages} {964} (\bibinfo {year} {1981})}\BibitemShut
  {NoStop}%
\bibitem [{\citenamefont {Caspers}\ \emph {et~al.}(1984)\citenamefont
  {Caspers}, \citenamefont {Emmett},\ and\ \citenamefont {Magnus}}]{caspe84}%
  \BibitemOpen
  \bibfield  {author} {\bibinfo {author} {\bibfnamefont {W.~J.}\ \bibnamefont
  {Caspers}}, \bibinfo {author} {\bibfnamefont {K.~M.}\ \bibnamefont {Emmett}},
  \ and\ \bibinfo {author} {\bibfnamefont {W.}~\bibnamefont {Magnus}},\
  }\href@noop {} {\bibfield  {journal} {\bibinfo  {journal} {J. Phys. A}\
  }\textbf {\bibinfo {volume} {17}},\ \bibinfo {pages} {2687} (\bibinfo {year}
  {1984})}\BibitemShut {NoStop}%
\bibitem [{\citenamefont {Brehmer}\ \emph {et~al.}(1998)\citenamefont
  {Brehmer}, \citenamefont {Kolezhuk}, \citenamefont {Mikeska},\ and\
  \citenamefont {Neugebauer}}]{Brehmer1998}%
  \BibitemOpen
  \bibfield  {author} {\bibinfo {author} {\bibfnamefont {S.}~\bibnamefont
  {Brehmer}}, \bibinfo {author} {\bibfnamefont {A.~K.}\ \bibnamefont
  {Kolezhuk}}, \bibinfo {author} {\bibfnamefont {H.-J.}\ \bibnamefont
  {Mikeska}}, \ and\ \bibinfo {author} {\bibfnamefont {U.}~\bibnamefont
  {Neugebauer}},\ }\href {http://stacks.iop.org/0953-8984/10/i=5/a=017}
  {\bibfield  {journal} {\bibinfo  {journal} {Journal of Physics: Condensed
  Matter}\ }\textbf {\bibinfo {volume} {10}},\ \bibinfo {pages} {1103}
  (\bibinfo {year} {1998})}\BibitemShut {NoStop}%
\bibitem [{\citenamefont {Tang}\ and\ \citenamefont
  {Sandvik}(2011)}]{Tang2011}%
  \BibitemOpen
  \bibfield  {author} {\bibinfo {author} {\bibfnamefont {Y.}~\bibnamefont
  {Tang}}\ and\ \bibinfo {author} {\bibfnamefont {A.~W.}\ \bibnamefont
  {Sandvik}},\ }\href {\doibase 10.1103/PhysRevLett.107.157201} {\bibfield
  {journal} {\bibinfo  {journal} {Phys. Rev. Lett.}\ }\textbf {\bibinfo
  {volume} {107}},\ \bibinfo {pages} {157201} (\bibinfo {year}
  {2011})}\BibitemShut {NoStop}%
\bibitem [{\citenamefont {Tang}\ and\ \citenamefont
  {Sandvik}(2015)}]{Tang2015}%
  \BibitemOpen
  \bibfield  {author} {\bibinfo {author} {\bibfnamefont {Y.}~\bibnamefont
  {Tang}}\ and\ \bibinfo {author} {\bibfnamefont {A.~W.}\ \bibnamefont
  {Sandvik}},\ }\href {\doibase 10.1103/PhysRevB.92.184425} {\bibfield
  {journal} {\bibinfo  {journal} {Phys. Rev. B}\ }\textbf {\bibinfo {volume}
  {92}},\ \bibinfo {pages} {184425} (\bibinfo {year} {2015})}\BibitemShut
  {NoStop}%
\bibitem [{\citenamefont {Dommange}\ \emph {et~al.}(2003)\citenamefont
  {Dommange}, \citenamefont {Mambrini}, \citenamefont {Normand},\ and\
  \citenamefont {Mila}}]{domma03}%
  \BibitemOpen
  \bibfield  {author} {\bibinfo {author} {\bibfnamefont {S.}~\bibnamefont
  {Dommange}}, \bibinfo {author} {\bibfnamefont {M.}~\bibnamefont {Mambrini}},
  \bibinfo {author} {\bibfnamefont {B.}~\bibnamefont {Normand}}, \ and\
  \bibinfo {author} {\bibfnamefont {F.}~\bibnamefont {Mila}},\ }\href@noop {}
  {\bibfield  {journal} {\bibinfo  {journal} {Phys. Rev. B}\ }\textbf {\bibinfo
  {volume} {68}},\ \bibinfo {pages} {224416} (\bibinfo {year}
  {2003})}\BibitemShut {NoStop}%
\bibitem [{\citenamefont {Balents}(2010)}]{balen10}%
  \BibitemOpen
  \bibfield  {author} {\bibinfo {author} {\bibfnamefont {L.}~\bibnamefont
  {Balents}},\ }\href@noop {} {\bibfield  {journal} {\bibinfo  {journal}
  {Nature}\ }\textbf {\bibinfo {volume} {464}},\ \bibinfo {pages} {199}
  (\bibinfo {year} {2010})}\BibitemShut {NoStop}%
\bibitem [{\citenamefont {Yan}\ \emph {et~al.}(2011)\citenamefont {Yan},
  \citenamefont {Huse},\ and\ \citenamefont {White}}]{yan11a}%
  \BibitemOpen
  \bibfield  {author} {\bibinfo {author} {\bibfnamefont {S.}~\bibnamefont
  {Yan}}, \bibinfo {author} {\bibfnamefont {D.~A.}\ \bibnamefont {Huse}}, \
  and\ \bibinfo {author} {\bibfnamefont {S.~R.}\ \bibnamefont {White}},\
  }\href@noop {} {\bibfield  {journal} {\bibinfo  {journal} {Science}\ }\textbf
  {\bibinfo {volume} {332}},\ \bibinfo {pages} {1173} (\bibinfo {year}
  {2011})}\BibitemShut {NoStop}%
\bibitem [{\citenamefont {Depenbrock}\ \emph {et~al.}(2012)\citenamefont
  {Depenbrock}, \citenamefont {McCulloch},\ and\ \citenamefont
  {Schollw\"ock}}]{depen12}%
  \BibitemOpen
  \bibfield  {author} {\bibinfo {author} {\bibfnamefont {S.}~\bibnamefont
  {Depenbrock}}, \bibinfo {author} {\bibfnamefont {I.~P.}\ \bibnamefont
  {McCulloch}}, \ and\ \bibinfo {author} {\bibfnamefont {U.}~\bibnamefont
  {Schollw\"ock}},\ }\href@noop {} {\bibfield  {journal} {\bibinfo  {journal}
  {Phys. Rev. Lett.}\ }\textbf {\bibinfo {volume} {109}},\ \bibinfo {pages}
  {067201} (\bibinfo {year} {2012})}\BibitemShut {NoStop}%
\bibitem [{\citenamefont {Norman}(2016)}]{norma16}%
  \BibitemOpen
  \bibfield  {author} {\bibinfo {author} {\bibfnamefont {M.~R.}\ \bibnamefont
  {Norman}},\ }\href@noop {} {\bibfield  {journal} {\bibinfo  {journal} {Rev.
  Mod. Phys.}\ }\textbf {\bibinfo {volume} {88}},\ \bibinfo {pages} {041002}
  (\bibinfo {year} {2016})}\BibitemShut {NoStop}%
\bibitem [{\citenamefont {Castelnovo}\ \emph {et~al.}(2008)\citenamefont
  {Castelnovo}, \citenamefont {Moessner},\ and\ \citenamefont
  {Sondhi}}]{caste08}%
  \BibitemOpen
  \bibfield  {author} {\bibinfo {author} {\bibfnamefont {C.}~\bibnamefont
  {Castelnovo}}, \bibinfo {author} {\bibfnamefont {R.}~\bibnamefont
  {Moessner}}, \ and\ \bibinfo {author} {\bibfnamefont {S.~L.}\ \bibnamefont
  {Sondhi}},\ }\href@noop {} {\bibfield  {journal} {\bibinfo  {journal}
  {Nature}\ }\textbf {\bibinfo {volume} {451}},\ \bibinfo {pages} {42}
  (\bibinfo {year} {2008})}\BibitemShut {NoStop}%
\bibitem [{\citenamefont {Morris}\ \emph {et~al.}(2009)\citenamefont {Morris},
  \citenamefont {Tennant}, \citenamefont {Grigera}, \citenamefont {Klemke},
  \citenamefont {Castelnovo}, \citenamefont {Moessner}, \citenamefont
  {Czternasty}, \citenamefont {Meissner}, \citenamefont {Rule}, \citenamefont
  {Hoffmann}, \citenamefont {Kiefer}, \citenamefont {Gerischer}, \citenamefont
  {Slobinsky},\ and\ \citenamefont {Perry}}]{morri09}%
  \BibitemOpen
  \bibfield  {author} {\bibinfo {author} {\bibfnamefont {D.~J.~P.}\
  \bibnamefont {Morris}}, \bibinfo {author} {\bibfnamefont {D.~A.}\
  \bibnamefont {Tennant}}, \bibinfo {author} {\bibfnamefont {S.~A.}\
  \bibnamefont {Grigera}}, \bibinfo {author} {\bibfnamefont {B.}~\bibnamefont
  {Klemke}}, \bibinfo {author} {\bibfnamefont {C.}~\bibnamefont {Castelnovo}},
  \bibinfo {author} {\bibfnamefont {R.}~\bibnamefont {Moessner}}, \bibinfo
  {author} {\bibfnamefont {C.}~\bibnamefont {Czternasty}}, \bibinfo {author}
  {\bibfnamefont {M.}~\bibnamefont {Meissner}}, \bibinfo {author}
  {\bibfnamefont {K.~C.}\ \bibnamefont {Rule}}, \bibinfo {author}
  {\bibfnamefont {J.-U.}\ \bibnamefont {Hoffmann}}, \bibinfo {author}
  {\bibfnamefont {K.}~\bibnamefont {Kiefer}}, \bibinfo {author} {\bibfnamefont
  {S.}~\bibnamefont {Gerischer}}, \bibinfo {author} {\bibfnamefont
  {D.}~\bibnamefont {Slobinsky}}, \ and\ \bibinfo {author} {\bibfnamefont
  {R.~S.}\ \bibnamefont {Perry}},\ }\href@noop {} {\bibfield  {journal}
  {\bibinfo  {journal} {Science}\ }\textbf {\bibinfo {volume} {326}},\ \bibinfo
  {pages} {411} (\bibinfo {year} {2009})}\BibitemShut {NoStop}%
\bibitem [{\citenamefont {Kadowaki}\ \emph {et~al.}(2009)\citenamefont
  {Kadowaki}, \citenamefont {Doi}, \citenamefont {Aoki}, \citenamefont
  {Tabata}, \citenamefont {Sato},\ and\ \citenamefont {Lynn}}]{kadow09}%
  \BibitemOpen
  \bibfield  {author} {\bibinfo {author} {\bibfnamefont {H.}~\bibnamefont
  {Kadowaki}}, \bibinfo {author} {\bibfnamefont {N.}~\bibnamefont {Doi}},
  \bibinfo {author} {\bibfnamefont {Y.}~\bibnamefont {Aoki}}, \bibinfo {author}
  {\bibfnamefont {Y.}~\bibnamefont {Tabata}}, \bibinfo {author} {\bibfnamefont
  {T.~J.}\ \bibnamefont {Sato}}, \ and\ \bibinfo {author} {\bibfnamefont
  {J.~W.}\ \bibnamefont {Lynn}},\ }\href@noop {} {\bibfield  {journal}
  {\bibinfo  {journal} {J. Phys. Soc. Jap.}\ }\textbf {\bibinfo {volume}
  {78}},\ \bibinfo {pages} {103706} (\bibinfo {year} {2009})}\BibitemShut
  {NoStop}%
\bibitem [{\citenamefont {Majumdar}\ and\ \citenamefont
  {Ghosh}(1969{\natexlab{a}})}]{Majumdar1969a}%
  \BibitemOpen
  \bibfield  {author} {\bibinfo {author} {\bibfnamefont {C.~K.}\ \bibnamefont
  {Majumdar}}\ and\ \bibinfo {author} {\bibfnamefont {D.~K.}\ \bibnamefont
  {Ghosh}},\ }\href {\doibase http://dx.doi.org/10.1063/1.1664978} {\bibfield
  {journal} {\bibinfo  {journal} {Journal of Mathematical Physics}\ }\textbf
  {\bibinfo {volume} {10}},\ \bibinfo {pages} {1388} (\bibinfo {year}
  {1969}{\natexlab{a}})}\BibitemShut {NoStop}%
\bibitem [{\citenamefont {Majumdar}\ and\ \citenamefont
  {Ghosh}(1969{\natexlab{b}})}]{Majumdar1969b}%
  \BibitemOpen
  \bibfield  {author} {\bibinfo {author} {\bibfnamefont {C.~K.}\ \bibnamefont
  {Majumdar}}\ and\ \bibinfo {author} {\bibfnamefont {D.~K.}\ \bibnamefont
  {Ghosh}},\ }\href {\doibase http://dx.doi.org/10.1063/1.1664979} {\bibfield
  {journal} {\bibinfo  {journal} {Journal of Mathematical Physics}\ }\textbf
  {\bibinfo {volume} {10}},\ \bibinfo {pages} {1399} (\bibinfo {year}
  {1969}{\natexlab{b}})}\BibitemShut {NoStop}%
\bibitem [{\citenamefont {Anderson}(1973)}]{Anderson1973}%
  \BibitemOpen
  \bibfield  {author} {\bibinfo {author} {\bibfnamefont {P.}~\bibnamefont
  {Anderson}},\ }\href {\doibase
  http://dx.doi.org/10.1016/0025-5408(73)90167-0} {\bibfield  {journal}
  {\bibinfo  {journal} {Materials Research Bulletin}\ }\textbf {\bibinfo
  {volume} {8}},\ \bibinfo {pages} {153 } (\bibinfo {year} {1973})}\BibitemShut
  {NoStop}%
\bibitem [{\citenamefont {Oguchi}\ and\ \citenamefont
  {Kitatani}(1989)}]{Oguchi1989}%
  \BibitemOpen
  \bibfield  {author} {\bibinfo {author} {\bibfnamefont {T.}~\bibnamefont
  {Oguchi}}\ and\ \bibinfo {author} {\bibfnamefont {H.}~\bibnamefont
  {Kitatani}},\ }\href {http://dx.doi.org/10.1143/JPSJ.58.1403} {\bibfield
  {journal} {\bibinfo  {journal} {J. Phys. Soc. Jap.}\ }\textbf {\bibinfo
  {volume} {58}},\ \bibinfo {pages} {1403} (\bibinfo {year}
  {1989})}\BibitemShut {NoStop}%
\bibitem [{\citenamefont {Liang}\ \emph {et~al.}(1988)\citenamefont {Liang},
  \citenamefont {Doucot},\ and\ \citenamefont {Anderson}}]{Liang1988}%
  \BibitemOpen
  \bibfield  {author} {\bibinfo {author} {\bibfnamefont {S.}~\bibnamefont
  {Liang}}, \bibinfo {author} {\bibfnamefont {B.}~\bibnamefont {Doucot}}, \
  and\ \bibinfo {author} {\bibfnamefont {P.~W.}\ \bibnamefont {Anderson}},\
  }\href {\doibase 10.1103/PhysRevLett.61.365} {\bibfield  {journal} {\bibinfo
  {journal} {Phys. Rev. Lett.}\ }\textbf {\bibinfo {volume} {61}},\ \bibinfo
  {pages} {365} (\bibinfo {year} {1988})}\BibitemShut {NoStop}%
\bibitem [{Note1()}]{Note1}%
  \BibitemOpen
  \bibinfo {note} {One can use the ``right'' and the ``left'' 1-spinon states
  also combined to construct orthonormal many-spinon states. For example, the
  2-spinon state $\mathinner {|{\Phi _{i,i+d}}\delimiter "526930B }$ with
  $d\geq 5$ can be constructed from the direct product of $\mathinner {|{\Phi
  _{i}^{\sigma ,\protect \text {r}}}\delimiter "526930B }$ and $\mathinner
  {|{\Phi _{i+d}^{\sigma ,\protect \text {l}}}\delimiter "526930B }$. However,
  it turns out that this leads to finite overlaps between different (for
  example 2- and 4-) spinon subspaces.}\BibitemShut {Stop}%
\bibitem [{\citenamefont {Wen}(1991)}]{wen91}%
  \BibitemOpen
  \bibfield  {author} {\bibinfo {author} {\bibfnamefont {X.~G.}\ \bibnamefont
  {Wen}},\ }\href@noop {} {\bibfield  {journal} {\bibinfo  {journal} {Phys.
  Rev. B}\ }\textbf {\bibinfo {volume} {44}},\ \bibinfo {pages} {2664}
  (\bibinfo {year} {1991})}\BibitemShut {NoStop}%
\bibitem [{\citenamefont {Okamoto}\ and\ \citenamefont
  {Nomura}(1992)}]{Okamoto1992}%
  \BibitemOpen
  \bibfield  {author} {\bibinfo {author} {\bibfnamefont {K.}~\bibnamefont
  {Okamoto}}\ and\ \bibinfo {author} {\bibfnamefont {K.}~\bibnamefont
  {Nomura}},\ }\href {\doibase http://dx.doi.org/10.1016/0375-9601(92)90823-5}
  {\bibfield  {journal} {\bibinfo  {journal} {Physics Letters A}\ }\textbf
  {\bibinfo {volume} {169}},\ \bibinfo {pages} {433 } (\bibinfo {year}
  {1992})}\BibitemShut {NoStop}%
\bibitem [{\citenamefont {Eggert}(1996)}]{Eggert1996}%
  \BibitemOpen
  \bibfield  {author} {\bibinfo {author} {\bibfnamefont {S.}~\bibnamefont
  {Eggert}},\ }\href {\doibase 10.1103/PhysRevB.54.R9612} {\bibfield  {journal}
  {\bibinfo  {journal} {Phys. Rev. B}\ }\textbf {\bibinfo {volume} {54}},\
  \bibinfo {pages} {R9612} (\bibinfo {year} {1996})}\BibitemShut {NoStop}%
\bibitem [{\citenamefont {Deschner}\ and\ \citenamefont
  {S\o{}rensen}(2013)}]{Deschner2013}%
  \BibitemOpen
  \bibfield  {author} {\bibinfo {author} {\bibfnamefont {A.}~\bibnamefont
  {Deschner}}\ and\ \bibinfo {author} {\bibfnamefont {E.~S.}\ \bibnamefont
  {S\o{}rensen}},\ }\href {\doibase 10.1103/PhysRevB.87.094415} {\bibfield
  {journal} {\bibinfo  {journal} {Phys. Rev. B}\ }\textbf {\bibinfo {volume}
  {87}},\ \bibinfo {pages} {094415} (\bibinfo {year} {2013})}\BibitemShut
  {NoStop}%
\bibitem [{\citenamefont {Bursill}\ \emph {et~al.}(1995)\citenamefont
  {Bursill}, \citenamefont {Gehring}, \citenamefont {Farnell}, \citenamefont
  {Parkinson}, \citenamefont {Xiang},\ and\ \citenamefont
  {Zeng}}]{Bursill1995}%
  \BibitemOpen
  \bibfield  {author} {\bibinfo {author} {\bibfnamefont {R.}~\bibnamefont
  {Bursill}}, \bibinfo {author} {\bibfnamefont {G.~A.}\ \bibnamefont
  {Gehring}}, \bibinfo {author} {\bibfnamefont {D.~J.~J.}\ \bibnamefont
  {Farnell}}, \bibinfo {author} {\bibfnamefont {J.~B.}\ \bibnamefont
  {Parkinson}}, \bibinfo {author} {\bibfnamefont {T.}~\bibnamefont {Xiang}}, \
  and\ \bibinfo {author} {\bibfnamefont {C.}~\bibnamefont {Zeng}},\ }\href
  {http://stacks.iop.org/0953-8984/7/i=45/a=016} {\bibfield  {journal}
  {\bibinfo  {journal} {Journal of Physics: Condensed Matter}\ }\textbf
  {\bibinfo {volume} {7}},\ \bibinfo {pages} {8605} (\bibinfo {year}
  {1995})}\BibitemShut {NoStop}%
\bibitem [{\citenamefont {Knetter}\ \emph {et~al.}(2003)\citenamefont
  {Knetter}, \citenamefont {Schmidt},\ and\ \citenamefont
  {Uhrig}}]{Knetter2003b}%
  \BibitemOpen
  \bibfield  {author} {\bibinfo {author} {\bibfnamefont {C.}~\bibnamefont
  {Knetter}}, \bibinfo {author} {\bibfnamefont {K.~P.}\ \bibnamefont
  {Schmidt}}, \ and\ \bibinfo {author} {\bibfnamefont {G.~S.}\ \bibnamefont
  {Uhrig}},\ }\href {http://stacks.iop.org/0305-4470/36/i=29/a=302} {\bibfield
  {journal} {\bibinfo  {journal} {J. Phys. A: Math. Gen.}\ }\textbf {\bibinfo
  {volume} {36}},\ \bibinfo {pages} {7889} (\bibinfo {year}
  {2003})}\BibitemShut {NoStop}%
\bibitem [{\citenamefont {Wegner}(1994)}]{Wegner1994}%
  \BibitemOpen
  \bibfield  {author} {\bibinfo {author} {\bibfnamefont {F.}~\bibnamefont
  {Wegner}},\ }\href {\doibase 10.1002/andp.19945060203} {\bibfield  {journal}
  {\bibinfo  {journal} {Annalen der Physik}\ }\textbf {\bibinfo {volume}
  {506}},\ \bibinfo {pages} {77} (\bibinfo {year} {1994})}\BibitemShut
  {NoStop}%
\bibitem [{\citenamefont {Kehrein}(2006)}]{Kehrein2006}%
  \BibitemOpen
  \bibfield  {author} {\bibinfo {author} {\bibfnamefont {S.}~\bibnamefont
  {Kehrein}},\ }\href
  {http://www.springer.com/materials/book/978-3-540-34067-6} {\emph {\bibinfo
  {title} {The Flow Equation Approach to Many-Particle Systems}}},\ Springer
  Tracts in Modern Physics, Vol. 217\ (\bibinfo  {publisher}
  {Berlin:Springer},\ \bibinfo {year} {2006})\BibitemShut {NoStop}%
\bibitem [{\citenamefont {Stein}(1997)}]{Stein1997}%
  \BibitemOpen
  \bibfield  {author} {\bibinfo {author} {\bibfnamefont {J.}~\bibnamefont
  {Stein}},\ }\href {\doibase 10.1007/BF02508481} {\bibfield  {journal}
  {\bibinfo  {journal} {Journal of Statistical Physics}\ }\textbf {\bibinfo
  {volume} {88}},\ \bibinfo {pages} {487} (\bibinfo {year} {1997})}\BibitemShut
  {NoStop}%
\bibitem [{\citenamefont {Mielke}(1998)}]{Mielke1998}%
  \BibitemOpen
  \bibfield  {author} {\bibinfo {author} {\bibfnamefont {A.}~\bibnamefont
  {Mielke}},\ }\href {\doibase 10.1007/s100510050485} {\bibfield  {journal}
  {\bibinfo  {journal} {Eur. Phys. J B}\ }\textbf {\bibinfo {volume} {5}},\
  \bibinfo {pages} {605} (\bibinfo {year} {1998})}\BibitemShut {NoStop}%
\bibitem [{\citenamefont {Fischer}\ \emph {et~al.}(2010)\citenamefont
  {Fischer}, \citenamefont {Duffe},\ and\ \citenamefont {Uhrig}}]{Fischer2010}%
  \BibitemOpen
  \bibfield  {author} {\bibinfo {author} {\bibfnamefont {T.}~\bibnamefont
  {Fischer}}, \bibinfo {author} {\bibfnamefont {S.}~\bibnamefont {Duffe}}, \
  and\ \bibinfo {author} {\bibfnamefont {G.~S.}\ \bibnamefont {Uhrig}},\ }\href
  {http://stacks.iop.org/1367-2630/12/i=3/a=033048} {\bibfield  {journal}
  {\bibinfo  {journal} {New Journal of Physics}\ }\textbf {\bibinfo {volume}
  {12}},\ \bibinfo {pages} {033048} (\bibinfo {year} {2010})}\BibitemShut
  {NoStop}%
\bibitem [{\citenamefont {Krull}\ \emph {et~al.}(2012)\citenamefont {Krull},
  \citenamefont {Drescher},\ and\ \citenamefont {Uhrig}}]{Krull2012}%
  \BibitemOpen
  \bibfield  {author} {\bibinfo {author} {\bibfnamefont {H.}~\bibnamefont
  {Krull}}, \bibinfo {author} {\bibfnamefont {N.~A.}\ \bibnamefont {Drescher}},
  \ and\ \bibinfo {author} {\bibfnamefont {G.~S.}\ \bibnamefont {Uhrig}},\
  }\href {\doibase 10.1103/PhysRevB.86.125113} {\bibfield  {journal} {\bibinfo
  {journal} {Phys. Rev. B}\ }\textbf {\bibinfo {volume} {86}},\ \bibinfo
  {pages} {125113} (\bibinfo {year} {2012})}\BibitemShut {NoStop}%
\bibitem [{\citenamefont {Yang}\ and\ \citenamefont
  {Schmidt}(2011)}]{Yang2011}%
  \BibitemOpen
  \bibfield  {author} {\bibinfo {author} {\bibfnamefont {H.~Y.}\ \bibnamefont
  {Yang}}\ and\ \bibinfo {author} {\bibfnamefont {K.~P.}\ \bibnamefont
  {Schmidt}},\ }\href {http://stacks.iop.org/0295-5075/94/i=1/a=17004}
  {\bibfield  {journal} {\bibinfo  {journal} {EPL (Europhysics Letters)}\
  }\textbf {\bibinfo {volume} {94}},\ \bibinfo {pages} {17004} (\bibinfo {year}
  {2011})}\BibitemShut {NoStop}%
\bibitem [{\citenamefont {Knetter}\ \emph {et~al.}(2000)\citenamefont
  {Knetter}, \citenamefont {B\"uhler}, \citenamefont {M\"uller-Hartmann},\ and\
  \citenamefont {Uhrig}}]{Knetter2000a}%
  \BibitemOpen
  \bibfield  {author} {\bibinfo {author} {\bibfnamefont {C.}~\bibnamefont
  {Knetter}}, \bibinfo {author} {\bibfnamefont {A.}~\bibnamefont {B\"uhler}},
  \bibinfo {author} {\bibfnamefont {E.}~\bibnamefont {M\"uller-Hartmann}}, \
  and\ \bibinfo {author} {\bibfnamefont {G.~S.}\ \bibnamefont {Uhrig}},\ }\href
  {\doibase 10.1103/PhysRevLett.85.3958} {\bibfield  {journal} {\bibinfo
  {journal} {Phys. Rev. Lett.}\ }\textbf {\bibinfo {volume} {85}},\ \bibinfo
  {pages} {3958} (\bibinfo {year} {2000})}\BibitemShut {NoStop}%
\bibitem [{\citenamefont {{Powalski}}\ \emph {et~al.}(2015)\citenamefont
  {{Powalski}}, \citenamefont {{Uhrig}},\ and\ \citenamefont
  {{Schmidt}}}]{Powalski2015}%
  \BibitemOpen
  \bibfield  {author} {\bibinfo {author} {\bibfnamefont {M.}~\bibnamefont
  {{Powalski}}}, \bibinfo {author} {\bibfnamefont {G.~S.}\ \bibnamefont
  {{Uhrig}}}, \ and\ \bibinfo {author} {\bibfnamefont {K.~P.}\ \bibnamefont
  {{Schmidt}}},\ }\href@noop {} {\bibfield  {journal} {\bibinfo  {journal}
  {Phys. Rev. Lett.}\ }\textbf {\bibinfo {volume} {115}},\ \bibinfo {pages}
  {207202} (\bibinfo {year} {2015})}\BibitemShut {NoStop}%
\bibitem [{\citenamefont {Heidbrink}\ and\ \citenamefont
  {Uhrig}(2002)}]{Heidbrink2002b}%
  \BibitemOpen
  \bibfield  {author} {\bibinfo {author} {\bibfnamefont {C.~P.}\ \bibnamefont
  {Heidbrink}}\ and\ \bibinfo {author} {\bibfnamefont {G.~S.}\ \bibnamefont
  {Uhrig}},\ }\href {\doibase 10.1103/PhysRevLett.88.146401} {\bibfield
  {journal} {\bibinfo  {journal} {Phys. Rev. Lett.}\ }\textbf {\bibinfo
  {volume} {88}},\ \bibinfo {pages} {146401} (\bibinfo {year}
  {2002})}\BibitemShut {NoStop}%
\bibitem [{\citenamefont {Hafez}\ and\ \citenamefont
  {Jafari}(2010)}]{Hafez2010b}%
  \BibitemOpen
  \bibfield  {author} {\bibinfo {author} {\bibfnamefont {M.}~\bibnamefont
  {Hafez}}\ and\ \bibinfo {author} {\bibfnamefont {S.~A.}\ \bibnamefont
  {Jafari}},\ }\href {\doibase 10.1140/epjb/e2010-10509-x} {\bibfield
  {journal} {\bibinfo  {journal} {The European Physical Journal B}\ }\textbf
  {\bibinfo {volume} {78}},\ \bibinfo {pages} {323} (\bibinfo {year}
  {2010})}\BibitemShut {NoStop}%
\bibitem [{\citenamefont {Kehrein}(1999)}]{kehre99}%
  \BibitemOpen
  \bibfield  {author} {\bibinfo {author} {\bibfnamefont {S.}~\bibnamefont
  {Kehrein}},\ }\href@noop {} {\bibfield  {journal} {\bibinfo  {journal} {Phys.
  Rev. Lett.}\ }\textbf {\bibinfo {volume} {83}},\ \bibinfo {pages} {4914}
  (\bibinfo {year} {1999})}\BibitemShut {NoStop}%
\bibitem [{\citenamefont {Lavar\'elo}\ and\ \citenamefont
  {Roux}(2014)}]{Lavarelo2014}%
  \BibitemOpen
  \bibfield  {author} {\bibinfo {author} {\bibfnamefont {A.}~\bibnamefont
  {Lavar\'elo}}\ and\ \bibinfo {author} {\bibfnamefont {G.}~\bibnamefont
  {Roux}},\ }\href {http://dx.doi.org/10.1140/epjb/e2014-50472-x} {\bibfield
  {journal} {\bibinfo  {journal} {The European Physical Journal B}\ }\textbf
  {\bibinfo {volume} {87}},\ \bibinfo {eid} {229} (\bibinfo {year}
  {2014})}\BibitemShut {NoStop}%
\bibitem [{\citenamefont {Chitra}\ \emph {et~al.}(1995)\citenamefont {Chitra},
  \citenamefont {Pati}, \citenamefont {Krishnamurthy}, \citenamefont {Sen},\
  and\ \citenamefont {Ramasesha}}]{Chitra1995}%
  \BibitemOpen
  \bibfield  {author} {\bibinfo {author} {\bibfnamefont {R.}~\bibnamefont
  {Chitra}}, \bibinfo {author} {\bibfnamefont {S.}~\bibnamefont {Pati}},
  \bibinfo {author} {\bibfnamefont {H.~R.}\ \bibnamefont {Krishnamurthy}},
  \bibinfo {author} {\bibfnamefont {D.}~\bibnamefont {Sen}}, \ and\ \bibinfo
  {author} {\bibfnamefont {S.}~\bibnamefont {Ramasesha}},\ }\href {\doibase
  10.1103/PhysRevB.52.6581} {\bibfield  {journal} {\bibinfo  {journal} {Phys.
  Rev. B}\ }\textbf {\bibinfo {volume} {52}},\ \bibinfo {pages} {6581}
  (\bibinfo {year} {1995})}\BibitemShut {NoStop}%
\bibitem [{\citenamefont {White}\ and\ \citenamefont
  {Affleck}(1996)}]{White1996}%
  \BibitemOpen
  \bibfield  {author} {\bibinfo {author} {\bibfnamefont {S.~R.}\ \bibnamefont
  {White}}\ and\ \bibinfo {author} {\bibfnamefont {I.}~\bibnamefont
  {Affleck}},\ }\href {\doibase 10.1103/PhysRevB.54.9862} {\bibfield  {journal}
  {\bibinfo  {journal} {Phys. Rev. B}\ }\textbf {\bibinfo {volume} {54}},\
  \bibinfo {pages} {9862} (\bibinfo {year} {1996})}\BibitemShut {NoStop}%
\bibitem [{\citenamefont {S\o{}rensen}\ \emph {et~al.}(1998)\citenamefont
  {S\o{}rensen}, \citenamefont {Affleck}, \citenamefont {Augier},\ and\
  \citenamefont {Poilblanc}}]{Sorensen1998}%
  \BibitemOpen
  \bibfield  {author} {\bibinfo {author} {\bibfnamefont {E.}~\bibnamefont
  {S\o{}rensen}}, \bibinfo {author} {\bibfnamefont {I.}~\bibnamefont
  {Affleck}}, \bibinfo {author} {\bibfnamefont {D.}~\bibnamefont {Augier}}, \
  and\ \bibinfo {author} {\bibfnamefont {D.}~\bibnamefont {Poilblanc}},\ }\href
  {\doibase 10.1103/PhysRevB.58.R14701} {\bibfield  {journal} {\bibinfo
  {journal} {Phys. Rev. B}\ }\textbf {\bibinfo {volume} {58}},\ \bibinfo
  {pages} {R14701} (\bibinfo {year} {1998})}\BibitemShut {NoStop}%
\end{thebibliography}
%

\end{document}